\newcommand{\ben}{\begin{equation}}
\newcommand{\een}{\end{equation}}
\newcommand{\bea}{\begin{eqnarray}}
\newcommand{\eea}{\end{eqnarray}}
\newcommand{\nn}{\nonumber\\ }
\newcommand{\rY}{{\rm Y}}
\newcommand{\rX}{{\rm X}}
\newcommand{\rZ}{{\rm Z}}
\newcommand{\rH}{{\rm H}}

\newcommand{\q}{{\qquad}}
\newcommand{\qq}{\qquad\qquad}
\newcommand{\QQ}{\qquad\qquad\qquad\qquad}
\newcommand{\ch}{{\rm ch}}
\newcommand{\hch}{\widehat{\rm ch}}
\newcommand{\hD}{\widehat{D}}
\newcommand{\Db}{{\bar{D}}}
\newcommand{\hd}{\hat{d}}

\newcommand{\hr}{\hat{r}}
\newcommand{\mult}{{\rm mult}}
\newcommand{\la}{\lambda}
\newcommand{\La}[1]{{\Lambda^{#1}}}
\newcommand{\vph}{\varphi}
\newcommand{\lf}{\lfloor}
\newcommand{\rf}{\rfloor}
\newcommand{\tWL}{{\overline{W\Lambda}}}
\newcommand{\cI}{{\cal I}}
\newcommand{\cO}{{\cal O}}
\input amssym.def
\def\Z{{\Bbb Z}} \def\R{{\Bbb R}} 
\def\C{{\Bbb C}} \def\O{{\Bbb O}} \def\H{{\Bbb H}}
\documentclass{article}
\usepackage{graphicx}
\usepackage{epsfig}
\begin{document}

\parskip=4pt
\baselineskip=14pt

\title{\vskip-1cm On character generators for simple Lie algebras}
\author{N. Okeke, M.A. Walton\\\\ {\it Department of
Physics,
University of Lethbridge}\\
{\em Lethbridge, Alberta, Canada\ \  T1K 3M4}\\\\
{\small nnamdi.okeke@uleth.ca, walton@uleth.ca}\\\\
}

\maketitle
\begin{abstract}
We study character generating functions (character generators) of
simple Lie algebras. The expression due to Patera and Sharp,
derived from the Weyl character formula, is first reviewed. A new
general formula is then found. It makes clear the distinct roles
of ``outside'' and ``inside'' elements of the integrity basis, and
helps determine their quadratic incompatibilities. We review,
analyze and extend the results obtained by Gaskell using the
Demazure character formulas. We find that the fundamental
generalized-poset graphs underlying the character generators can
be deduced from such calculations. These graphs, introduced by
Baclawski and Towber, can be simplified for the purposes of
constructing the character generator. The generating functions can
be written easily using the simplified versions, and associated
Demazure expressions. The rank-two algebras are treated in detail,
but we believe our results are indicative of those for general
simple Lie algebras.
\end{abstract}

\vfill\eject
\section{Introduction}

Characters are important tools in the representation theory of Lie
groups and algebras, and so are relevant to many physical
applications. Generating function techniques are powerful and
general, and have been usefully applied in many areas of
mathematics and mathematical physics.

Combining characters and generating functions leads to the study
of character generators \cite{PS}, the generating functions of
characters. We will report results on the character generators for
the irreducible, integrable, highest-weight representations of
simple Lie algebras.

Much work has been done on these character generators, and on
other generating functions relevant to Lie algebras (see \cite{P}
for a brief summary). Our emphasis will be on general results,
valid for all simple Lie algebras and expressed in terms of
structures common to them.\footnote{\ Such relations are sometimes
called universal.} Relevant highlights of past research include
the papers by Patera and Sharp \cite{PS}, Stanley \cite{S}, King
\cite{K}, Baclawski \cite{Ba}, King and El-Sharkaway
\cite{KEi,KEii}, and Baclawski and Towber \cite{BT}.

Patera and Sharp \cite{PS} introduced the character generator, and
used the Weyl character formula to write a general formula for it.
While it is general, the Patera-Sharp formula has the same
drawbacks as its summand, the Weyl character formula. It involves
positive and negative terms that cancel, and a sum over the Weyl
group. The Weyl group sum is enormous for all but the lowest rank
algebras. In the first of two parts of this paper, starting from
the Patera-Sharp formula, we will derive a general formula for the
character generator that doesn't involve that sum.

The first drawback still applies to the new formula, however. To
write non-negative formulas, a more microscopic point of view
helps. The character generator is a generating function for
characters, but characters are themselves generating functions for
weight multiplicities. Combinatorial methods for the calculation
of these multiplicities are well known. In particular, methods
involving Young tableaux and variants are very efficient. Stanley
\cite{S} was the first to apply them to the character generator.
In particular, he proved a formula for the $A_r\cong su(r+1)$
character generator involving standard shifted Young tableaux.
Soon afterward, King \cite{K} adapted Stanley's work to the case
$C_r\cong sp(2r)$.

Most importantly for us, these standard shifted Young tableaux
encode a partially ordered set, or poset. The poset structure of
the character generator was made explicit by Baclawski \cite{Ba}.
He proved combinatorial formulas for the character generators of
$A_r$, $C_r$ (and $U(N)$) in terms of poset objects related to the
relevant tableaux.

King and El-Sharkaway generalized these results to all the
classical algebras in \cite{KEi,KEii}, using generalized standard
Young tableaux. This work is very, but not completely, general:
besides the classical simple Lie algebras $A_r,B_r,C_r,D_r$, there
are also the exceptional ones $E_6,E_7,E_8,F_4,G_2$. The notion of
standard Young tableau was not general enough to include the
exceptional Lie algebras.\footnote{\ More recently, however,
Littelmann has provided such generalizations, in the form of
minimal defining chains of elements of the Weyl group \cite{L},
and also in terms of so-called Lakshmibai-Seshadri paths and root
operators \cite{Li}.}

Baclawski and Towber \cite{BT} studied the simplest exceptional
algebra, $G_2$. They were able to construct the character
generator, and found that a generalization of a poset was
relevant. This generalized-poset structure is important, but it
was revealed by a construction specific to $G_2$, related to the
octonions. A truly general construction was therefore not found.

Meanwhile, however, new general methods were applied to the
problem. The Demazure character formulas \cite{D} are valid for
all simple Lie algebras (and others), and lead directly to
non-negative expressions for the characters. Gaskell \cite{G}
wrote a general formula for the character generator as a Demazure
operator acting on the highest-weight generating function (see
(\ref{Gi}) below), and calculated several examples.\footnote{\
Actually, Gaskell was unaware of Demazure's work. Remarkably, he
re-discovered some of the Demazure formulas independently, in
order to apply them to character generators.}

In the second part of this paper, we will push Gaskell's methods,
and apply Demazure character formulas to the character generator.
Most importantly, we will also search for the underlying
generalized-poset structure of the character generator. By
combining the Gaskell-Demazure techniques with the
generalized-poset structure, we make progress. While our
calculations focus on the simple Lie algebras of rank two, we
believe that our results indicate those for all simple Lie
algebras. We are able to conjecture a general non-negative formula
for the character generator of an arbitrary simple Lie algebra.

The layout of the paper is as follows. In the next section, we
first review the derivation of the Patera-Sharp formula from the
Weyl character formula, and then derive a new general formula from
the Patera-Sharp one. It makes clear the distinct roles of the
inside and outside generators of the integrity basis for the terms
of the character generator. As we show in Sect. 3, it can also
serve as a guide to the incompatible products (consequences of
syzygies) of the elements of the integrity basis. In Sect. 4, we
review the Demazure character formulas, and apply them to the
construction of character generators, following Gaskell. Simple,
non-negative expressions are found for the character generators of
all the rank-two simple Lie algebras. In Sect. 5, the poset and
generalized-poset structures of character generators are reviewed,
and then applied to our rank-two results. We simplify the
generalized-poset graphs introduced by Baclawski and Towber, and
introduce edge labels for the new graphs. These edge labels are
expressed in terms of Demazure operators, and so can be simply
determined. Consequently, we are able to write a formula that is
valid for all the rank-two algebras, and others, in terms of a
so-called fundamental-orbit poset and Demazure quantities. We
conjecture that this formula is universal, i.e. applicable to all
simple Lie algebras. Sect. 6 is our conclusion.

\vskip1cm\section{Character generators from the Weyl character
formula}

Let ${\rm X}(L,a)$ denote the generator (generating function) for
the characters of a fixed simple Lie algebra $X_r$, of rank $r$.
It is defined \cite{PS} \ben {\rm X}(L,a)\ :=\ \sum_{\lambda\in
P_\ge}\, L^\lambda\,\ch_\la(a) \ \ , \label{defX} \een where the
character of the integrable, irreducible representation $R(\la)$
of highest weight $\la$ is \ben \ch_\lambda(a)\ =\ \sum_{\sigma\in
P}\, \mult_\lambda(\sigma)\,a^\sigma\ \ .\label{defch}\een Two
sets of indeterminate variables are used. We write
 \ben L^\lambda\ =\ L^{\sum_i\lambda_i\Lambda^i}\ :=\ L_1^{\la_1}\cdots
L_r^{\la_r}\ \ \label{Lla}\een to keep track of the highest
weights of representations, and $a^\mu := a_1^{\mu_1}\cdots
a_r^{\mu_r}$ to record the weights with nonvanishing
multiplicities in those representations. In (\ref{defX}),
$\mult_\la(\sigma)$ is the multiplicity of weight $\sigma$ in
$R(\la)$.

The fundamental weights are the $\Lambda^j$, and the set thereof
will be denoted $F$. The set of integral weights of $X_r$ is \ben
P\ :=\ \left\{ \sum_{i=1}^r\,\lambda_i\Lambda^i \,|\, \la_i\in{\Z}
\right\}\ ,\label{Pdef}\een  i.e., the set of weights with integer
Dynkin labels $\la_i$. $P_{\ge}\subset P$ will be the set of
dominant weights \ben P_\ge\ :=\ \left\{
\sum_{i=1}^r\,\lambda_i\Lambda^i \,|\, \la_i\in{\Z_\ge} \right\}\
,\label{Pge}\een with non-negative integer (semi-natural) Dynkin
labels. Similarly, $P_{>}\subset P$ will denote the set of weights
\ben P_>\ :=\ \left\{ \sum_{i=1}^r\,\lambda_i\Lambda^i \,|\,
\la_i\in{\Z_>} \right\}\ ,\label{Pgt}\een with positive integer
(natural) Dynkin labels. We will also sometimes use the notation
\ben \la\ =\ \sum_{i=1}^r\,\la_i \Lambda^i\ =:\
(\la_1,\la_2,\ldots,\la_r)\ .\label{vec}\een The set of weights of
representation $R(\lambda)$ will be indicated by \ben P_\lambda\
:=\ \left\{\ \mu\in P \ |\ \mult_\la(\mu)\ge 1 \ \right\}\ \
.\label{Pldef}\een

The Weyl formula for the character of $R(\lambda)$ is \bea
\ch_\la(a)\ =\ &\sum_{w\in W}\, a^{w\lambda}\,
\prod_{\alpha\in\Delta_+}\, (1-a^{-w\alpha})^{-1}\ \nn =\
&\prod_{\alpha\in\Delta_+}\, (1-a^{-\alpha})^{-1}\,\sum_{w\in W}\,
(\det w)\, a^{w.\lambda}\ .\label{wcf}\eea $W$ is the Weyl group
of $X_r$, $\Delta_+$ the set of its positive roots, and
$w.\la=w(\la+\rho)-\rho$ is the shifted action of the Weyl group
element $w\in W$. The Weyl vector is denoted \ben \rho\ =\
\frac{1}{2}\sum_{\alpha\in\Delta_+}\,\alpha\ =\
\Lambda^1+\Lambda^2+\ldots+\Lambda^r\ =\ \sum_{\Lambda\in
F}\,\Lambda\ .\label{wvec}\een

The key observation of \cite{PS} is that if the Weyl character
formula (\ref{wcf}) is used, the sum over $P_\ge$ in (\ref{defX})
can be done, yielding the Patera-Sharp formula \ben {\rm X}(L,a)\
=\ \prod_{\alpha\in\Delta_+}\,(1-a^{-\alpha})^{-1}\,\sum_{w\in
W}\,a^{w\rho-\rho}\,(\det w)\, \prod_{\Lambda\in F}\,
\left(1-L^\Lambda a^{w\Lambda}
 \right)^{-1}\ . \label{PSf}\een This is already a nice,
general result. However, the sum over the Weyl group is daunting
for any but the smallest Lie algebras. Furthermore, division by
the Weyl denominator
$\prod_{\alpha\in\Delta_+}\,(1-a^{-\alpha})^{-1}$ makes direct
computation quite difficult.

However, if we factor out a common denominator, call it ${\rm Z}$,
things improve somewhat. As is usual, let $W\lambda$ indicate the
set of weights in the Weyl orbit of $\lambda$. Then we can write
\ben {\rm X}(L,a)\ =\ \prod_{\Lambda\in F}\,\prod_{\vph\in
W\Lambda}\, \left(1-L^\Lambda a^{\vph}
 \right)^{-1}\ {\rm Y}\ =:\ {\rm Z}^{-1}\,\,{\rm Y}\ ,\label{XYZ}\een
 with \bea {\rm Y}\ =\ \prod_{\alpha\in\Delta_+}\,
(1-a^{-\alpha})^{-1}\QQ\QQ\nn \quad\times\,\sum_{w\in
W}\,a^{w\rho-\rho}\,(\det w)\, \prod_{\Lambda\in
F}\,\,\prod_{\sigma\in W\Lambda\backslash\{w\Lambda\}}\,\, \left(
1-L^\Lambda a^\sigma \right)\ \ .\label{iYi}\eea

It is well known that the characters may be written as integer
polynomials of the fundamental characters. We therefore expect
that the integrity basis $I_\rX$ will be \ben I_\rX\ =\ \left\{\,
L^\Lambda a^\vph\, |\, \Lambda\in F,\ \vph\in P_\Lambda
\,\right\}\ .\label{IX}\een That is, we expect that the character
generator $\rX$ can be written as a rational function of the
elements of $I_\rX$. For the integrity-basis element $L^\Lambda
a^\vph$, $\Lambda$ and $\varphi$ will be known as its shape (or
its highest weight) and its weight, respectively.

Clearly, it is the numerator ${\rm Y}$ that encodes the truly
nontrivial information carried by a character generator ${\rm X}$.
The denominator ${\rm Z}$ tells us only that the ``outside
weights'' of the fundamental representations determine a subset
\ben I_{\rm out}\ =\ \left\{\, L^\Lambda a^\vph\, |\, \Lambda\in
F,\ \vph\in W\Lambda \,\right\}\  \label{Xbas}\een of the
integrity basis $I_\rX$ for the terms of ${\rm X}$. The elements
of $I_{\rm out}$ and $I_{\rm in}:= I_\rX\backslash I_{\rm out}$
will be called outside and inside generators, respectively.

Another helpful observation is that the terms in the sum of
(\ref{iYi}) are simply related to each other. Let $\hat w$ denote
an ``operator'' with action \ben \hat w\,\left(\, L^\nu
a^\mu\,\right)\ :=\ L^\nu a^{w\mu}\ ,\label{hatw}\een for any
weights $\nu, \mu$. Then we can write \ben {\rm Y}\  =\
\prod_{\alpha\in\Delta_+}\,(1-a^{-\alpha})^{-1}\, \sum_{w\in
W}\,a^{w\rho-\rho}\,(\det w)\,\hat w\,\left(\, {\cal Y}\,\right)\
,\label{Yii}\een where we have defined \ben {\cal Y}\ :=\
\prod_{\Lambda\in F}\,\prod_{\sigma\in \tWL}\, \left( 1-L^\Lambda
a^\sigma \right) \label{Yi}\een  and the shorthand \ben \tWL\ :=\
W\Lambda\backslash\{\Lambda\}\ .\label{WLudef}\een Now, comparing
with the Weyl formula (\ref{wcf}), we see that \ben
\prod_{\alpha\in\Delta_+}\,(1-a^{-\alpha})^{-1}\,\sum_{w\in
W}\,a^{w\rho-\rho}\,(\det w)\,\hat w\ =:\ \hat{\ch}\
\label{hatch}\een acts as follows: \ben \hat{\ch}\,\left(\,
a^\lambda \,\right)\ =\
\prod_{\alpha\in\Delta_+}\,(1-a^{-\alpha})^{-1}\,\sum_{w\in
W}\,a^{w\rho-\rho}\,(\det w)\,\hat w\, \left(\, a^\lambda
\,\right)\ =\ \ch_\la(a)\ .\label{hcha}\een Therefore, we get \ben
{\rm Y}\ =\ \hch\ \left(\, {\cal Y} \,\right)\ =\ {\hat{\ch}}\,
\left(\  \prod_{\Lambda\in F}\,\prod_{\sigma\in \tWL}\, \left(
1-L^\Lambda a^\sigma \right)\ \right)\ .\label{Yiii}\een

This formula shows that we can decompose $\rm Y$ into characters,
\ben {\rm Y}\ =\ \sum_{\mu\in P_{\ge}}\, {\rm y}_\mu(L)\,\ch_\mu\
.\label{dY}\een The coefficients ${\rm y}_\mu(L)$ will be
polynomials in the $L_j=L^{\Lambda^j}$, with integer coefficients.
To evaluate ${\rm Y}$ in this form, we use the shifted-Weyl
(anti-)symmetry of the characters \ben \ch_\la\ =\ (\det
w)\,\ch_{w.\la}\ .\label{swch}\een

If we define a partition function $K_\mu(L)$ as
follows:\footnote{\ Here we imitate the definition of the Kostant
partition function. See Sect. 25.2 of \cite{FH}, for example.}
\ben {\mathcal{Y}}(L,a)\ =\ \prod_{\Lambda\in F}\,\prod_{\sigma\in
\tWL}\, \left( 1-L^\Lambda a^\sigma \right)\ =:\ \sum_{\tau\in
P}\, K_\tau(L)\, a^\tau\ ,\label{K}\een then the desired
coefficients can be computed using \ben {\rm y}_\mu(L)\ =\
\sum_{w\in W}\, (\det w)\, K_{w.\mu}(L)\ \ .\label{ywK}\een This
equation says that the ${\rm y}_\mu$ can be calculated by first
expanding ${\mathcal{Y}}$, using its definition (see eqn.
(\ref{Yi})). Each term obtained with $a$-dependence $a^\varphi$
can be Weyl-transformed using the shifted action, so that the
result $a^\nu$ has $\nu+\rho$ either in $P_>$, or on its boundary
(i.e. having at least one vanishing Dynkin label) . In the latter
case, the term should be dropped. In the former case, it
contributes with an extra factor of $\det w$, where $w$ is the
Weyl group element used. All terms $a^\nu$ so collected, with
$\nu\in P_\ge$, signal a contribution of $\ch_\nu$ to ${\rm Y}$.
We hope that the examples worked through in the following section
will make the procedure clear.

Formally, then, the answer is \ben {\rm Y}(L,a)\ =\ \sum_{\nu\in
P_\ge}\,\ch_\nu(a)\, \sum_{w\in W}\, (\det w)\, K_{w.\nu}(L)\ \ ,
\label{YywK}\een so that the character generator is \ben {\rm
X}(L,a)\, =\, \bigg\{\, \prod_{{\Lambda\in F}\atop{\vph\in
W\Lambda}}\, \left(1-L^\Lambda a^{\vph}
 \right) \,\bigg\}^{-1}\, \sum_{\nu\in P_\ge}\,\ch_\nu(a) \sum_{w\in
W}\, (\det w)\, K_{w.\nu}(L)\ . \label{XZwK}\een

\vskip.5cm\subsection{Examples}

\subsubsection{$A_1$}

For $X_r=A_1$, there is only one fundamental weight, $\Lambda^1$,
and we have \ben {\rm Z}\ =\ (1-L^{\Lambda^1}a^{\Lambda^1})
(1-L^{\Lambda^1}a^{-\Lambda^1})\ \ . \label{ZAi} \een Since ${\cal
Y}\ =\ (1-L^{\Lambda^1}a^{-\Lambda^1})$, we have \ben {\rm Y}\ =\
1-L^{\Lambda^1}\ch_{-\Lambda^1}\ ,\label{YAi}\een by (\ref{Yiii}).
But $\ch_{-\Lambda^1}=-\ch_{-\Lambda^1}=0$, by (\ref{swch}), so
that ${\rm Y}=1$. Finally, we have the well-known result \ben {\rm
X}(L,a)\ =\ \left[ (1-L^{\Lambda^1 }a^{\Lambda^1}) (1-L^{\Lambda^1
}a^{-\Lambda^1}) \right]^{-1}\ .\label{XAi}\een

\subsubsection{$A_2$}

\bea {\rm Z}\ =\ &(1-L^{\Lambda^1}a^{\Lambda^1})
(1-L^{\Lambda^1}a^{-\Lambda^1+\Lambda^2})(1-L^{\Lambda^1}a^{-\Lambda^2})
\qq\nn & \qq\times \,(1-L^{\Lambda^2}a^{\Lambda^2})
(1-L^{\Lambda^2}a^{\Lambda^1-\Lambda^2})(1-L^{\Lambda^2}a^{-\Lambda^1})\
\ , \label{ZAii} \eea and \ben {\cal Y}\ =\
(1-L^{\Lambda^1}a^{-\Lambda^1+\Lambda^2})(1-L^{\Lambda^1}a^{-\Lambda^2})
(1-L^{\Lambda^2}a^{\Lambda^1-\Lambda^2})(1-L^{\Lambda^2}a^{-\Lambda^1})\
.\label{YiAii}\een From (\ref{Yiii}), expanding ${\cal Y}$ and
applying $\hch$ gives \bea &{\rm Y}\ =\
1-L^{\Lambda^1}(\ch_{-\Lambda^1+\Lambda^2}+\ch_{-\Lambda^2}) -
L^{\Lambda^2}(\ch_{\Lambda^1-\Lambda^2}+\ch_{-\Lambda^1})\nn &+
L^{2\Lambda^1}\ch_{-\Lambda^2} +
L^{2\Lambda^2}\ch_{-\Lambda^2}\qq\nn
&+L^{\Lambda^1+\Lambda^2}(1+\ch_{-2\Lambda^1+\Lambda^2}+\ch_{\Lambda^1-2\Lambda^2}
+\ch_{-\Lambda^1-\Lambda^2}) \nn
&-L^{2\Lambda^1+\Lambda^2}(\ch_{-\Lambda^2}+\ch_{-2\Lambda^1})
-L^{\Lambda^1+2\Lambda^2}(\ch_{-\Lambda^1}+\ch_{-2\Lambda^2})\nn
&+L^{2\Lambda^1+2\Lambda^2}\ch_{-\Lambda^1-\Lambda^2}\
.\label{YchiAii}\eea Any term $\ch_\mu$ with a Dynkin label
$\mu_i=-1$ vanishes, since if $r_i$ denotes the primitive
reflection related to the simple root $\alpha_i$, then
$r_i.\mu=\mu$. Eqn. (\ref{swch}) then tells us that
$\ch_\mu=-\ch_\mu=0$. The expression immediately simplifies to
\bea {\rm Y}\ =\
1+L^{\Lambda^1+\Lambda^2}(1+\ch_{-2\Lambda^1+\Lambda^2}+\ch_{\Lambda^1-2\Lambda^2})\nn
-L^{2\Lambda^1+\Lambda^2}\ch_{-2\Lambda^1}
-L^{\Lambda^1+2\Lambda^2}\ch_{-2\Lambda^2}\ .\label{YchiiAii}\eea
But $r_1.(-2\Lambda^1)=-\Lambda^2$ and
$r_2.(-2\Lambda^2)=-\Lambda^1$, so the last 2 terms vanish. Also,
$r_1.(-2\Lambda^1+\Lambda^2)=r_2.(\Lambda^1-2\Lambda^2)=0$, so
that we obtain \ben {\rm Y}\ =\ 1-L^{\Lambda^1+\Lambda^2}\ \ .
\label{YchiiiAii}\een Finally, we can write \ben {\rm X}\ =\ {\rm
Z}^{-1}\, \left[1-L^{\Lambda^1+\Lambda^2}\right]\ ,
\label{XAii}\een with ${\rm Z}$ given by (\ref{ZAii}), in
agreement with the known result \cite{PS}.

\subsubsection{$B_2$}

The simple roots are $\alpha_1=2\Lambda^1-2\Lambda^2$ and
$\alpha_2=-\Lambda^1+2\Lambda^2$. The Weyl orbits of the
fundamental weights, \bea W\Lambda^1\ =\ \left\{\ \pm\Lambda^1,
\pm(-\Lambda^1+2\Lambda^2) \right\}\ \nn W\Lambda^2\ =\ \left\{\
\pm\Lambda^2, \pm(\Lambda^1-\Lambda^2) \right\}\ ,\
\label{WLa}\eea determine both ${\rm Z}$, by (\ref{XYZ}), and
${\rm Y}$, by (\ref{Yiii}).

For convenience, we write explicitly \bea {\rm Z}\ =\
(1-L^{\Lambda^1}a^{\Lambda^1})
(1-L^{\Lambda^1}a^{-\Lambda^1+2\Lambda^2})
(1-L^{\Lambda^1}a^{\Lambda^1-2\Lambda^2}) \nn \times \
(1-L^{\Lambda^1}a^{-\Lambda^1}) (1-L^{\Lambda^2}a^{\Lambda^2})
(1-L^{\Lambda^2}a^{\Lambda^1-\Lambda^2}) \nn \times \
(1-L^{\Lambda^2}a^{-\Lambda^1+\Lambda^2})
(1-L^{\Lambda^2}a^{-\Lambda^2})\ , \label{ZBii}\eea and \bea {\rm
Y}\ =\ \hch\ (1-L^{\Lambda^1}a^{-\Lambda^1+2\Lambda^2})
(1-L^{\Lambda^1}a^{\Lambda^1-2\Lambda^2})
(1-L^{\Lambda^1}a^{-\Lambda^1}) \nn \times \
(1-L^{\Lambda^2}a^{\Lambda^1-\Lambda^2})
(1-L^{\Lambda^2}a^{-\Lambda^1+\Lambda^2})
(1-L^{\Lambda^2}a^{-\Lambda^2})\ .\qq \label{YBii}\eea As a first
step in evaluating $\rY$, we expand the last expression, dropping
any characters with Dynkin labels equalling -1. The result is \bea
{\rm Y} {\hskip-.5cm}&=\ \hch\ \big\{\, 1-L^\La1 a^{\La1-2\La2} +
L^{2\La1}(1+a^{-2\La2}+a^{-2\La1+2\La2})\qq\nn
&+L^{\La1+\La2}(a^{-2\La1+\La2}+a^{-2\La1+3\La2}+
a^{2\La1-3\La2}+a^{\La2}+a^{\La1-3\La2})\nn
&+L^{2\La2}(1+a^{\La1-2\La2})
-L^{2\La1+\La2}(a^{-3\La2}+a^{-3\La1+3\La2}+a^{\La1-3\La2}\nn
&+a^{-2\La1+\La2}) -L^{\La1+2\La2}(1+a^{\La1-2\La2}\nn &
+a^{-2\La1}+a^{-2\La1+2\La2} +a^{2\La1-4\La2}+2a^{-2\La2})
+L^{3\La1+\La2}a^{-2\La1+\La2} \nn &+ L^{2\La1+2\La2}
(a^{-2\La2}+a^{-2\La1+2\La2}+1 +a^{\La1-4\La2} +a^{\La1-2\La2}
+a^{-3\La1+2\La2})\nn & +L^{\La1+3\La2}a^{\La1-3\La2}
-L^{3\La1+2\La2}(a^{-2\La1}+a^{-2\La2})\nn &
-L^{2\La1+3\La2}(a^{-3\La2}+ a^{-2\La1+\La2})\,\big\}\
.\label{YiBii}\eea Not only will the contributions of weights
$(-1, \mu_2)$ and $(\mu_1, -1)$ vanish, but so will those of any
in their shifted Weyl orbits, $W.(-1, \mu_2)$: \ben \{ (-1,\mu_2),
(\mu_2,-\mu_2-2), (-\mu_2-2,\mu_2), (-1,-\mu_2-2) \}\
,\label{Wdmi}\een and  $W.(\mu_1, -1)$: \ben  \{ (\mu_1,-1),
(-\mu_1-2,2\mu_1+1), (\mu_1,-2\mu_1-3), (-\mu_2-1,-1) \}\
.\label{Wdmii}\een We can therefore also eliminate any terms
$a^\mu$ with $\mu$ of forms $(a,-a-2)$ and $(a,-2a-3)$. Doing
this, we find \bea {\rm Y} \ =\ \hch\ \big\{\, 1-L^\La1
a^{\La1-2\La2} + L^{2\La1}(1+a^{-2\La1+2\La2}) \qq\nn
+L^{\La1+\La2}(a^{-2\La1+3\La2}+ a^{2\La1-3\La2}+a^{\La2}) \qq\nn
+L^{2\La2}(1+a^{\La1-2\La2}) -L^{2\La1+\La2} a^{-3\La1+3\La2}
\qq\nn -L^{\La1+2\La2}(1+a^{\La1-2\La2} +a^{-2\La1+2\La2} ) \qq\ \
\nn + L^{2\La1+2\La2} (a^{-2\La1+2\La2}+1 +a^{\La1-4\La2}
+a^{\La1-2\La2} +a^{-3\La1+2\La2}) \,\}\ . \label{YiiBii}\eea
Calculating \bea  r_1.(-2,3)\ =\ r_2.(2,-3)\ =\ (0,1)\ ,\qq\qq\nn
r_2.(1,-2) = r_1.(-2,2) = (r_1r_2).(1,-4) = (r_2r_1).(-3,2) =\ 0\
\ ,
 \label{Wdmisc}\eea we find \ben {\rm Y}\ =\ 1+L^\La1
  - L^{\La1+\La2} \ch_\La2 + L^{\La1+2\La2} +
 L^{2\La1+2\La2}\  .\label{YiiBiii}\een  Using  \ben  \ch_\La2\
 =\ a^\La2+a^{-\La1+\La2}
 +a^{-\La2}+a^{\La1-\La2}\ ,\label{chii}\een
 we have checked that this answer agrees with the known result \cite{PS}.

\subsubsection{$G_2$}

 The simple roots are $\alpha_{1}=2\Lambda^{1}-3\Lambda^{2}$
 and $\alpha_{2}=-\Lambda^{1}+2\Lambda^{2}$. The Weyl orbits of the
fundamental weights are:
\begin{eqnarray}
W\Lambda^{1}=\{\pm\Lambda^{1},\pm(\Lambda^{1}-3\Lambda^{2}),\pm(2\Lambda^{1}-
3\Lambda^2)\}\ ,
\nonumber\\
W\Lambda^{2}=\{\pm\Lambda^{2},\pm(\Lambda^{1}-\Lambda^{2}),
\pm(\Lambda^{1}-2\Lambda^2)\}\
.
\end{eqnarray}
Therefore,
\begin{eqnarray}
Z\ =\
(1-L^{\Lambda^{1}}a^{\Lambda^{1}})(1-L^{\Lambda^{1}}a^{\Lambda^{1}-3\Lambda^{2}})
(1-L^{\Lambda^{1}}a^{-\Lambda^{1}+3\Lambda^{2}})\nonumber\\
\times\,
(1-L^{\Lambda^{1}}a^{2\Lambda^{1}-3\Lambda^{2}})(1-L^{\Lambda^{1}}a^{-2\Lambda^{1}+3\Lambda^{2}})
(1-L^{\Lambda^{1}}a^{-\Lambda^{1}})\nonumber\\ \times\,
(1-L^{\Lambda^{2}}a^{\Lambda^{2}})(1-L^{\Lambda^{2}}a^{\Lambda^{1}-\Lambda^{2}})
(1-L^{\Lambda^{2}}a^{\Lambda^{1}-2\Lambda^{2}})\nonumber\\
\times\, (1-L^{\Lambda^{2}}a^{-\Lambda^{1}+\Lambda^{2}})
(1-L^{\Lambda^{2}}a^{-\Lambda^{1}+2\Lambda^{2}})(1-L^{\Lambda^{2}}a^{-\Lambda^{2}})
\end{eqnarray}
and
\begin{eqnarray}
{\cal Y}=(1-L^{\Lambda^{1}}a^{\Lambda^{1}-3\Lambda^{2}})
(1-L^{\Lambda^{1}}a^{-\Lambda^{1}+3\Lambda^{2}})
(1-L^{\Lambda^{1}}a^{2\Lambda^{1}-3\Lambda^{2}})\nonumber \\
\times(1-L^{\Lambda^{1}}a^{-2\Lambda^{1}+3\Lambda^{2}})
(1-L^{\Lambda^{1}}a^{-\Lambda^{1}})(1-L^{\Lambda^{2}}a^{\Lambda^{1}-\Lambda^{2}})\nonumber
\\
\times(1-L^{\Lambda^{2}}a^{\Lambda^{1}-2\Lambda^{2}})(1-L^{\Lambda^{2}}a^{-\Lambda^{1}+\Lambda^{2}})
(1-L^{\Lambda^{2}}a^{-\Lambda^{1}+2\Lambda^{2}})(1-L^{\Lambda^{2}}a^{-\Lambda^{2}})
\end{eqnarray}
By (\ref{Yiii}), expanding ${\cal Y}$ and applying $\hch$ gives
\bea Y \,= \, 1 + L^{\Lambda^{1}}\, +\, L^{\Lambda^{2}} \, +\,
L^{3\Lambda^{1}+3\Lambda^{2}}\,+\,L^{\Lambda^{1}+4\Lambda^{2}}\qq\nn
\, +\, L^{3\Lambda^{1}} \,+\,L^{4\Lambda^{1}+3\Lambda^{2}} \,+\,
L^{\Lambda^{1}+\Lambda^{2}}\,+\,L^{\Lambda^{1}+3\Lambda^{2}} \nn
\, +\, L^{\Lambda^{1}+2\Lambda^{2}} \,+\,
L^{3\Lambda^{1}+2\Lambda^{2}} \, + \,
L^{4\Lambda^{1}+4\Lambda^{2}}  \,+\, L^{2\Lambda^{1}+4\Lambda^{2}}
\nn \,+\, L^{3\Lambda^{1}+4\Lambda^{2}} \,+\,
L^{3\Lambda^{1}+\Lambda^{2}} \,+\, L^{2\Lambda^{1}} \,+\,
(L^{3\Lambda^{1}+4\Lambda^{2}} \,+\, L^{2\Lambda^{1}+4\Lambda^{2}}
\q\nn \qq\,+\, L^{\Lambda^{1}} \,+ \,L^{2\Lambda^{1}}  \,+\,
2L^{2\Lambda^{1}+2\Lambda^{2}}\, -\, L^{3\Lambda^{1}+\Lambda^{2}}
\, -\, L^{\Lambda^{1}+3\Lambda^{2}}) \,\ch_{\Lambda^{2}} \nn \,+\,
(L^{3\Lambda^{1}+2\Lambda^{2}} \,+\, L^{\Lambda^{1}+2\Lambda^{2}}
\, -\, L^{2\Lambda^{1}+\Lambda^{2}} \,-\,
L^{2\Lambda^{1}+3\Lambda^{2}}) \,\ch_{\Lambda^{1}} \nn \,+\,
L^{2\Lambda^{1}+2\Lambda^{2}} \,\ch_{\Lambda^{1}+\Lambda^{2}}\QQ
\nn \,-\, ( L^{\Lambda^{1}+\Lambda^{2}}  \,+\,
L^{2\Lambda^{1}+\Lambda^{2}} \, +\, L^{3\Lambda^{1}+3\Lambda^{2}}
\, +\, L^{2\Lambda^{1}+3\Lambda^{2}}) \,\ch_{2\Lambda^{2}}
 \eea
after applying (\ref{swch}). Using the following characters
\begin{eqnarray}
\ch_{\Lambda^{1}}=2+a^{\Lambda^{1}}+a^{-\Lambda^{1} +3\Lambda^{2}}
+a^{2\Lambda^{1}-3\Lambda^{2}}
+a^{-2\Lambda^{1}+3\Lambda^{2}}+a^{\Lambda^{1}-3\Lambda^{2}}\nonumber\\+a^{-\Lambda^{1}}+a^{\Lambda^{2}}
+a^{-\Lambda^{1}+2\Lambda^{2}}+a^{-\Lambda^{1}+\Lambda^{2}}+a^{-\Lambda^{2}}\nonumber\\+
a^{\Lambda^{1}-2\Lambda^{2}} +a^{\Lambda^{1}-\Lambda^{2}}\ ,
\end{eqnarray}
\begin{eqnarray}
\ch_{\Lambda^{2}}=1+a^{\Lambda^{2}}+a^{\Lambda^{1}-\Lambda^{2}}+a^{-\Lambda^{1}+2\Lambda^{2}}\nonumber\\
+a^{\Lambda^{1}-2\Lambda^{2}}+a^{-\Lambda^{1}+\Lambda^{2}}+a^{-\Lambda^{2}}\
,
\end{eqnarray}
\begin{eqnarray}
\ch_{\Lambda^{1}+\Lambda^{2}}=4+2a^{2\Lambda^{2}}+4a^{\Lambda^{2}}+2a^{\Lambda^{1}}+
2a^{2\Lambda^{1}-3\Lambda^{2}}\nonumber\\
+2a^{-\Lambda^{1}+3\Lambda^{2}}+2a^{-2\Lambda^{1}+3\Lambda^{2}}+2a^{\Lambda^{1}-
3\Lambda^{2}}+2a^{-\Lambda^{1}}\nonumber\\
+4a^{-\Lambda^{1}+2\Lambda^{2}}+4a^{-\Lambda^{1}+\Lambda^{2}}+4a^{\Lambda^{2}}+
4a^{\Lambda^{1}-2\Lambda^{2}}\nonumber\\
+4a^{\Lambda^{1}-\Lambda^{2}}+2a^{2\Lambda^{1}-2\Lambda^{2}}+2a^{2\Lambda^{1}-
4\Lambda^{2}}+2a^{-2\Lambda^{2}}\nonumber\\
+2a^{-2\Lambda^{1}+2\Lambda^{2}}+2a^{-2\Lambda^{1}+4\Lambda^{2}}+
a^{\Lambda^{1}+\Lambda^{2}}\nonumber\\
+a^{2\Lambda^{1}-\Lambda^{2}}+a^{3\Lambda^{1}-5\Lambda^{2}}+ a^{2\Lambda^{1}-5\Lambda^{2}}\nonumber\\
+a^{\Lambda^{1}-4\Lambda^{2}}+a^{-2\Lambda^{1}+\Lambda^{2}}+a^{-3\Lambda^{1}+4\Lambda^{2}}\nonumber\\
+a^{-3\Lambda^{1}+5\Lambda^{2}}+a^{-2\Lambda^{1}+5\Lambda^{2}}\nonumber\\
+a^{-\Lambda^{1}+4\Lambda^{2}}+a^{-\Lambda^{1}-\Lambda^{2}}+a^{3\Lambda^{1}-4\Lambda^{2}}\ ,\nonumber\\
\end{eqnarray}
and
\begin{eqnarray}
\ch_{2\Lambda^{2}}=3+2a^{\Lambda^{2}}+2a^{\Lambda^{1}-\Lambda^{2}}+
2a^{-\Lambda^{1}+2\Lambda^{2}}\nonumber\\
+2a^{\Lambda^{1}-2\Lambda^{2}}+2a^{-\Lambda^{1}+\Lambda^{2}}+
2a^{-\Lambda^{2}}+a^{\Lambda^{1}}\nonumber\\
+a^{-\Lambda^{1}+3\Lambda^{2}}+a^{2\Lambda^{1}-3\Lambda^{2}}
+a^{-2\Lambda^{1}+3\Lambda^{2}}\nonumber\\
+a^{\Lambda^{1}-3\Lambda^{2}}+a^{-\Lambda^{1}}+a^{2\Lambda^{1}-2\Lambda^{2}}\nonumber\\
+a^{2\Lambda^{1}-4\Lambda^{2}}+a^{-2\Lambda^{2}}+a^{-2\Lambda^{1}+2\Lambda^{2}}\nonumber\\
+a^{-2\Lambda^{1}+4\Lambda^{2}} +a^{2\Lambda^{2}}\ ,
\end{eqnarray}
we have verified that this expression agrees with the known result
\cite{GS}.

\vskip1cm\section{Integrity basis and incompatibilities}

The characters can be generated by an integrity basis subject to
certain relations. The basis is given in (\ref{IX}). For a fixed
simple Lie algebra, the character of any irreducible
representation can be written as an non-negative integer
polynomial in these basis elements.

Important relations can be expressed as incompatibilities,
quadratic products of basis elements that do not appear in any of
the monomials just mentioned. Here we show how the new formula
(\ref{XZwK}) can be used as a guide to the incompatibilities. A
different method was described in \cite{HS}.

Since the outside and inside generators ($I_{\rm in} :=
I_\rX\backslash I_{\rm out}$) play different roles in generating
characters, it will be useful to split the fundamental characters
into inside and outside parts, by writing \ben \ch_\Lambda(a)\ =:\
{\cal O}_\Lambda(a)\ +\ {\cal I}_\Lambda(a)\ .\label{ioch}\een
Here we denote the orbit sum by \ben {\cal O}_\lambda(a) \ =\
\sum_{\sigma\in W\lambda}\, a^{\sigma}\  .\label{orb}\een

The numerator $\rY$ contains the required information. Using
(\ref{Yiii}), we will expand $\rY$ up to terms quadratic in the
$L_i$: \ben \rY\ =\ \rY^{(0)}+ \rY^{(1)}+ \rY^{(2)}+ \ldots\ =\
\hch\, \big(\, {\cal Y}^{(0)}+ {\cal Y}^{(1)}+ {\cal Y}^{(2)}+
\ldots \,\big)\ . \label{Yex}\een

Clearly, \ben \rY^{(0)}\ =\ {\cal Y}^{(0)}\ =\ 1\ . \label{Yz}\een
The linear term is \ben \rY^{(1)}\ =\ -\hch\,\left(
\sum_{\Lambda\in F}\ \sum_{\phi\in \tWL}\,  L^\Lambda a^\phi
\right)\ =\ \hch\,\left( \sum_{\Lambda\in F}\, L^\Lambda\, \big(\,
a^\Lambda- {\cal O}_\Lambda(a) \,\big) \right)\ ; \label{Yexi}\een
see (\ref{orb}). We now use the identity \ben \hch\, \left(\,
a^\mu \, {\cal O}_\lambda(a) \,\right)\ =\ \ch_\mu(a)\, {\cal
O}_\lambda(a)\ ,\label{chO}\een proved in the Appendix, with
$\mu=0$. We find \bea \rY^{(1)}\ =&\ \sum_{\Lambda\in F}\,
L^\Lambda\, \big(\, \ch_\Lambda(a) - {\cal O}_\Lambda(a)
  \,\big)\ \nn =:&\ \sum_{\Lambda\in F}\, L^\Lambda\,
  {\cal I}_\Lambda\ .\qq\qq\label{Yexpi}\eea The result
  shows explicitly that the inside
generators all appear linearly in $\rX$: \ben \rY^{(1)}\ =\
\sum_{\iota\in I_{\rm in}}\, \iota\ ,\label{Yexai}\een for all
simple Lie algebras.

The quadratic term can be expressed as \ben {\cal Y}^{(2)}\ =\
\sum_{\Lambda\in F}\ L^{2\Lambda}
\sum_{{\phi,\phi'\in\tWL}\atop{\phi'\not=\phi}}\, a^{\phi+\phi'}\
+\ \sum_{{\Lambda,\Lambda'\in F}\atop{\Lambda\not=\Lambda'}}\,
L^{\Lambda+\Lambda'}
\sum_{{\phi\in\tWL}\atop{\phi'\in{\overline{W\Lambda'}}}}\
a^{\phi+\phi'}\ .\label{Yippii}\een This leads to the expression
\bea  \rY^{(2)}\ =\ \sum_{{\Lambda,\Lambda'\in
F}\atop{\Lambda'\not=\Lambda}}\, \Big(\, \cI_\Lambda\cI_{\Lambda'}
- S_{\Lambda,\Lambda'} \,\Big)\, L^{\Lambda+\Lambda'}\ \qq\nn \qq
+\ \sum_{\Lambda\in F}\, \Big(\,  \ch_{2\Lambda}
-S_{\Lambda,\Lambda} + \cI_\Lambda^{\,2} - \cO_{2\Lambda}
\,\Big)\, L^{2\Lambda}\ .
 \label{YiIO}\eea
Here we have defined \ben S_{\Lambda,\Lambda'}\ :=\
\ch_\Lambda\ch_{\Lambda'}\ -\ \ch_{\Lambda+\Lambda'}\ .
\label{Sdef}\een

Similar expressions for terms $\rY^{(n)}$ with $n>2$ are
complicated. We will focus on the quadratic term $\rY^{(2)}$
below, and treat each of the rank-two simple Lie algebras in turn.

\vskip.5cm\subsection{Examples}

\subsubsection{$A_2$}

From subsection 2.1.2, \ben \rY^{(2)}\ =\  -L_1L_2\ .
\label{AiiYq}\een This result agrees with that calculated using
(\ref{Yippii}).

For any algebra $A_r$, $\cI_\Lambda=0$ for all $\Lambda\in F$.
Using \bea \ch_{\Lambda^1}\,\ch_{\Lambda^2}\ =\
\ch_{\Lambda^1+\Lambda^2}\ +\ 1\ ,\qq\qq\nn ( \ch_{\Lambda^1} )^2\
=\ \ch_{2\Lambda^1}\
 +\ \ch_{\Lambda^2}\ ,\ \ ( \ch_{\Lambda^2}
)^2\ =\ \ch_{2\Lambda^2}\
 +\ \ch_{\Lambda^1}\ , \nn \ch_{2\Lambda^1}\ -\
 \cO_{2\Lambda^1}\ =\ \ch_{\Lambda^2}\ ,\ \ \ch_{2\Lambda^2}\ -\
 \cO_{2\Lambda^2}\ =\ \ch_{\Lambda^1}\ ,\ \ \ \label{chidsAii}\eea
 (\ref{YiIO}) gives the same result.

 The interpretation of the result (\ref{AiiYq}) is simple. There
 is one incompatible quadratic product, but it is not
 uniquely determined. Any of the 3 following possibilities works:
 \ben \,(L_1a_1)(L_2a_1^{-1})\, ,\ \ (L_1a_1^{-1}a_2)(L_2a_1a_2^{-1})\,
 ,\ \ (L_1a_2^{-1})(L_2a_2)\, .\ \label{fpAii}\een We will see
 below in subsect. 4.1.1 that these choices lead to three different,
 but equivalent, expressions for $\rX$.

\subsubsection{$B_2$}

To use (\ref{YiIO}), we need \bea \cI_{\Lambda^1}\ =\ 1\, ,\ \
\cI_{\Lambda^2}\ =\ 0\ ,\qq\qq\nn S_{\Lambda^1,\Lambda^1}\ =\ 1\
+\ \ch_{2\Lambda^2}\ ,\ \ S_{\Lambda^1,\Lambda^2}\ =\
\ch_{\Lambda^2}\ ,\ \ S_{\Lambda^2,\Lambda^2}\ =\ 1\ +\
\ch_{\Lambda^1}\ ,\nn \ch_{2\Lambda^1}\ -\ \cO_{2\Lambda^1}\ =\
\ch_{2\Lambda^2}\ ,\ \ \ch_{2\Lambda^2}\ -\ \cO_{2\Lambda^2}\ =\
1\ +\  \ch_{\Lambda^1}\ ,\label{chidsBii}\eea to find  \ben
\rY^{(2)}\ =\ -L_1L_2\,\ch_{\Lambda^2}\ . \label{BiiYq}\een This
is in  agreement with the result of subsection 2.1.3.

The negative terms in (\ref{BiiYq}) reveal incompatibilities
between generators. One choice of incompatible products is \bea
(L_1a_1)(L_2a_1^{-1}a_2),\ (L_1a_1)(L_2a_2^{-1}),\ \nn
(L_1a_1^{-1}a_2^2)(L_2a_2^{-1}),\ (L_1)(L_2a_2^{-1})\
.\label{fpBii}\eea The sum of these four terms equals
$L_1L_2\,\ch_{\Lambda^2}$, therefore agreeing with (\ref{BiiYq}).

In subsection 4.1.2 below we will relate this choice of
incompatible products to a non-negative expression for $\rX$, and
an underlying graph.

\subsubsection{$G_2$}

From subsection 2.1.4, \ben \rY^{(2)}\ =\  L_1^2\,\big(
1+\ch_{\Lambda^2} \big)\ +\ L_1L_2\,\big( 1-\ch_{2\Lambda^2}
\big)\ , \label{GiiYq}\een in accord with that calculated using
(\ref{Yippii}).

To verify (\ref{YiIO}), \bea \cI_{\Lambda^1}\ =\ 1\ +\
\ch_{\Lambda^2}\ ,\ \ \ \cI_{\Lambda^2}\ =\ \ch_{\Lambda^1}\ +\
\ch_{\Lambda^2}\ ,\qq\nn S_{\Lambda^1,\Lambda^2}\ =\ 1\ +\
\ch_{\Lambda^1}\ +\ \ch_{2\Lambda^2}\ +\ \ch_{3\Lambda^2}\ ,\qq\nn
S_{\Lambda^1,\Lambda^2}\ =\ \ch_{\Lambda^2}\ +\ \ch_{2\Lambda^2}\
,\ \ S_{\Lambda^2,\Lambda^2}\ =\ 1\ +\ \ch_{\Lambda^1}\ +\
\ch_{\Lambda^2}\ , \nn \ch_{2\Lambda^1}\ -\ \cO_{2\Lambda^1}\ =\
\ch_{3\Lambda^2}\ -\ch_{\Lambda^2}\ +\ 1\ ,\qq\nn
\ch_{2\Lambda^2}\ -\ \cO_{2\Lambda^2}\ =\ \ch_{\Lambda^1}\ +\
\ch_{\Lambda^2}\ ,\qq\label{chidsGii}\eea are useful.

We will verify in subsect. 4.1.3 below that the expression
(\ref{GiiYq}) encodes the incompatible products for $\rX$. More
precisely, we will show that it can be written as a sum of terms
\ben \rY^{(2)}\ =\ -\ \rY^{(2)}_{\rm out,out}\ +\ \rY^{(2)}_{\rm
in,in}\ -\ \rY^{(2)}_{\rm in,out}\ .\label{inoutYiiGii}\een The
negative terms are incompatible products, either with two outer
generators as factors, or one inner and one outer. Since the
factor $Z^{-1}$ of $\rX$ does not involve the inner generators,
the allowed products quadratic in the inner generators appear in
$\rY^{(2)}$; that explains the positive term.

\vskip1cm\section{Gaskell character generators from Demazure
character formulas}

In this section, we follow Gaskell \cite{Ga} and apply the
Demazure character formulas to the calculation of character
generators. We will be able to interpret our results in terms of
certain graphs, as discussed in Sect. 5.

Let us first review the Demazure character formula(s), and set our
notation. Demazure \cite{D} introduced the operators $\hD_i,\
i=1,\ldots,r$, associated with the simple roots of the Lie algebra
$X_r$, or the corresponding primitive reflections $r_i$. They are
defined by the action \ben \hD_i\, (\, a^\phi \,)\ =\
\left\{\matrix{ a^{\phi} +a^{\phi -\alpha_i}+ \ldots
+a^{\phi-\phi_i\alpha_i}\ ,\ &\phi_i\ge 0\ ; \cr \qq \ \ 0\ ,\
\qq\ \ \ \ \ &\phi_i=-1\ ; \cr -a^{\phi+\alpha_i}
-a^{\phi+2\alpha_i} - \ldots -a^{\phi+(|\phi_i|-1)\alpha_i}\ ,\ \
&\phi_i\le-2\ .} \right. \label{Diap}\een The number of terms in
these expansions is $|\phi_i+1|$. Alternatively, one can write
\ben \hD_i\ =\ (1-a^{-\alpha_i})^{-1}\,(1-a^{-\alpha_i}\hr_i)\ .
\label{Dia}\een

A unique Demazure operator can be defined for every element of the
Weyl group $W$. Suppose $w\in W$ has a reduced decomposition
$w=s_\ell\cdots s_2s_1$. Here each $s_j=r_{j'}$ is a primitive
reflection. Since the decomposition is reduced, $\ell$ is the
minimum possible length, the length $\ell(w)$ of $w$. Then we can
define \ben \hD_w\ :=\ \hD_{1'}\,\hD_{2'}\cdots \hD_{\ell'}\
.\label{hDw}\een Reduced decompositions are not unique, however.
For example, the longest element $w_L$ of the $su(3)$ Weyl group
$W\cong S_3$ has two such decompositions, \ben w_L\ =\ r_1r_2r_1\
=\ r_2r_1r_2\ .\label{suiiibraid}\een But the braid relation that
equates them is also satisfied by the Demazure operators: \ben
\hD_{w_L}\ =\ \hD_1\hD_2\hD_1\ =\ \hD_2\hD_1\hD_2\
,\label{Dsuiiibraid}\een so that $\hD_{w_L}$ can be constructed
using either of its reduced decompositions. Such braid relations
are obeyed for any simple Lie algebra, and the operators $\hD_w$
are uniquely defined for any $w\in W$. The basic operators are the
$\hD_i:=\hD_{r_i}.$

Notice, however that although the braid relations of the Weyl
group are obeyed by the Demazure operators, we have $r_i^2=1$, but
$\hD_i^2\not= 1$. Instead \ben \big( \hD_i \big)^2\ =\ \hD_i\ ,
\label{hDii}\een so that the $\hD_i$ are projection operators. It
is also very useful to realize that \ben \hD_i\,(1+\hr_i)\ =\
(1+\hr_i)\ ,\label{Diinv}\een so that $\hD_i$ does not change
expressions that are $\hr_i$-invariant. Using this fact can reduce
computations significantly.

The Demazure character formula can be written simply as \ben
\ch_\lambda(a)\ =\ \hD_L\, \left(\, a^\lambda \,\right) ,
\label{DemDL}\een where we have written $\hD_L:= \hD_{w_L}$ for
short. Equivalently, we can write \ben \hch\ =\ \hD_L\ ,
\label{hchDL}\een for the operator $\hch$ introduced in
(\ref{hatch}).

As an example, consider the $su(3)$ representation of highest
weight $\lambda=2\Lambda^1+\Lambda^2$. We will use the reduced
decomposition $\hD_L = \hD_1\hD_2\hD_1$. First, \ben \hD_1\,
a^{2\Lambda^1+\Lambda^2}\ =\ a^{2\Lambda^1+\Lambda^2} +
a^{2\Lambda^2} + a^{-2\Lambda^1+3\Lambda^2}\ .\label{DemDLxi}\een
Then \bea  \hD_2\,\hD_1\,  a^{2\Lambda^1+\Lambda^2}\ =\ \big(\,
a^{2\Lambda^1+\Lambda^2} + a^{3\Lambda^1-\Lambda^2} \,\big) \QQ\nn
+ \big(\, a^{2\Lambda^2} + a^{\Lambda^1} +
a^{2\Lambda^1-2\Lambda^2} \,\big) \qq\nn\qq + \big(\,
a^{-2\Lambda^1+3\Lambda^2} + a^{-\Lambda^1+ \Lambda^2} +
a^{-\Lambda^2} + a^{\Lambda^1-3\Lambda^2} \,\big)\ .
\label{DemDLxii}\eea To avoid generating terms with negative
integer coefficients, that will eventually cancel anyway, we
separate out the $\hr_1$-invariant part of this result, \ben
\big(\, a^{2\Lambda^1+\Lambda^2} + a^{2\Lambda^2} +
a^{-2\Lambda^1+3\Lambda^2} \,\big) + \big(\, a^{\Lambda^1} +
a^{-\Lambda^1+ \Lambda^2} \,\big) + \big(\, a^{-\Lambda^2} \,\big)
\label{riinvt}\een before applying $\hD_1$. By virtue of
(\ref{Diinv}), we then need only compute \bea \hD_1\, \big(\,
a^{3\Lambda^1-\Lambda^2} + a^{2\Lambda^1-2\Lambda^2} +
a^{\Lambda^1-3\Lambda^2} \,\big)\  =\ \QQ\nn\qq \big(\,
a^{3\Lambda^1-\Lambda^2} + a^{\Lambda^1} +
a^{-\Lambda^1+\Lambda^2} + a^{-3\Lambda^1+2\Lambda^2} \,\big)\ +\
\qq\nn  \quad \big(\, a^{2\Lambda^1-2\Lambda^2} + a^{-\Lambda^2} +
a^{-2\Lambda^1} \,\big) +\ \big(\, a^{\Lambda^1-3\Lambda^2} +
a^{-\Lambda^1-2\Lambda^2} \,\big)\ . \label{DemDLxiii}\eea Adding
this last result to (\ref{riinvt}) then gives the character
$\ch_{2\Lambda^1+\Lambda^2}(a)$, without the need for
cancellations between positive and negative terms, as in the Weyl
character formula.

Also useful are operators $\hd_i$, defined by \ben
 \hD_i\ =:\ 1+\hd_i\ ,\ \ \hd_i\ :=\
 (1-a^{-\alpha_i})^{-1}\,a^{-\alpha_i}\, (1-\hr_i)\ .\label{Demd}\een Their
 action is \ben  \hd_i\, (\, a^\phi \,)\ =\ \left\{\matrix{ a^{\phi-\alpha_i}
+a^{\phi -2\alpha_i}+ \ldots +a^{\phi-\phi_i\alpha_i}\ ,\
&\phi_i\ge 1\ ; \cr \qq \ \ 0\ ,\ \qq\ \ \ \ \ &\phi_i=0\ ; \cr
-a^\phi-a^{\phi+\alpha_i} - \ldots -a^{\phi+(|\phi_i|-1)\alpha_i}\
,\ \ &\phi_i\le-1\ .} \right. \label{diap}\een Notice that the
number of terms in all three of these last expressions is
$|\phi_i|$.

\begin{figure}[t]
\begin{minipage}{10cm}
\vskip-0cm
\begin{center}
\epsfxsize=11cm \epsfbox{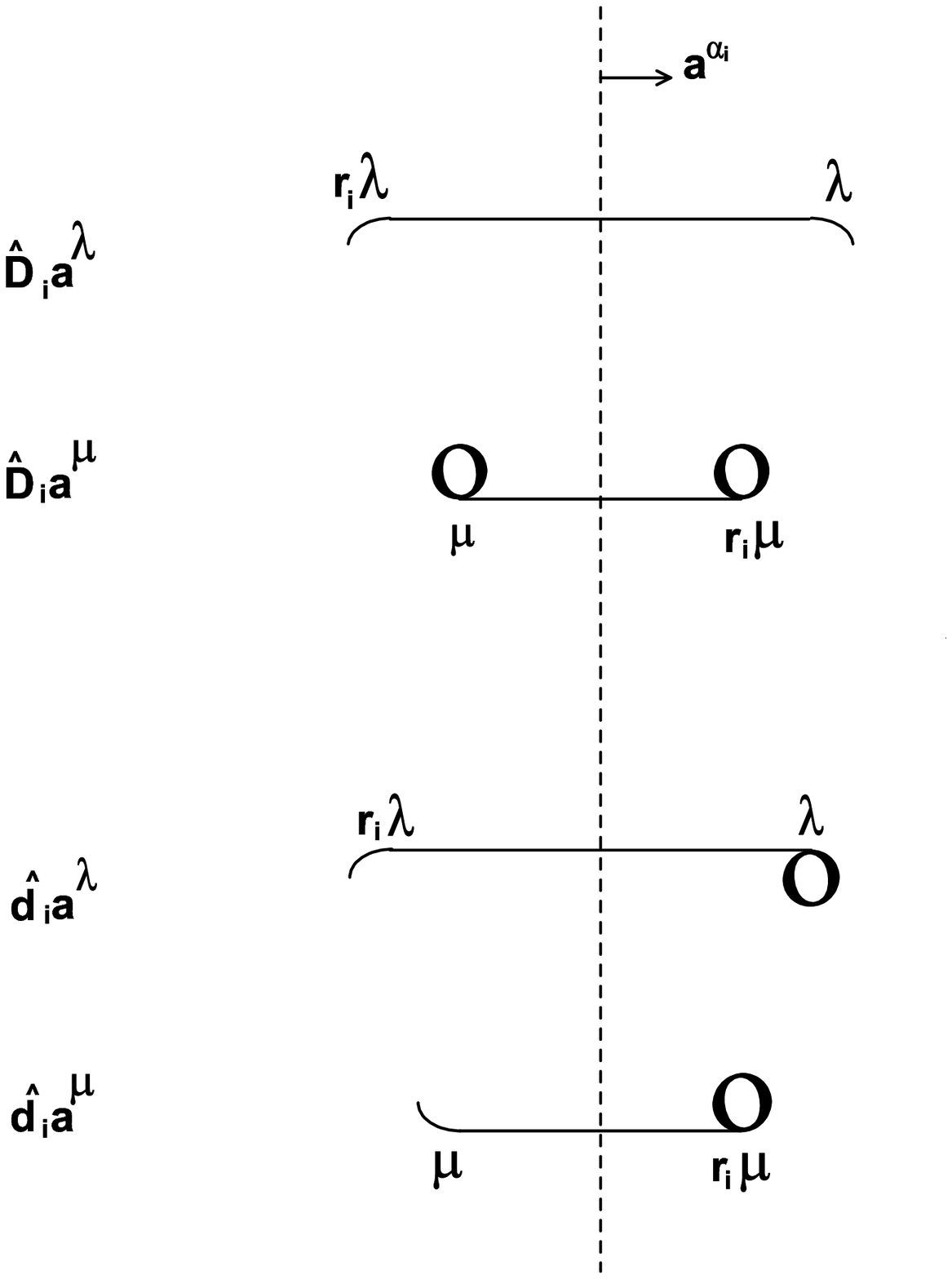} \vskip-4.5cm \caption{\quad Action
of the Demazure operators $\hD_i$ and $\hd_i$. The weight $\la$ has
positive, integer Dynkin label $\la_i$, while $\mu_i$ is a negative
integer.}
\label{Dd} 
\end{center}
\end{minipage}
\end{figure}

Cartoons of the actions of the Demazure operators are given in
Fig. \ref{Dd}. They make clear certain relations, such as
$\hr_i\hD_i = \hD_i$, $\hD_i = \hd_i+1$, $\hr_i\hd_i =
a^{\alpha_i}\hd_i$, $\hd_i\hr_i = -\hd_i$, etc. The vertical,
dashed line in the figure represents the hyperplane in weight
space where the $i$-th Dynkin label vanishes. The actions are
indicated both for a weight $\lambda$, with positive Dynkin label
$\lambda_i$, and a weight $\mu$, with $\mu_i<0$. Raised,
horizontal lines represent strings of terms like $a^\lambda+
a^{\lambda-\alpha_i} + ... + a^{r_i\lambda}$, with positive
coefficients +1. Lowered, horizontal lines correspond to such
strings with -1 as their coefficients. The circles, consisting as
they do of a raised and a lowered part, contribute 0, but
emphasize that there is no term $a^\lambda$ in $\hd_ia^\lambda$,
e.g.

A unique operator $\hd_w$ can again be defined for any $w\in W$,
using  reduced decompositions of $w$, if we set \ben \hd_{id}\,
\left(\, a^\lambda \,\right)\ =\ a^\lambda\ . \label{hdid}\een In
agreement with (\ref{hDii}), we have \ben \hd_i^{\ 2}\ +\ \hd_i\
=\ 0\ ,\label{hdii}\een so that the Demazure character formula
(\ref{DemDL}) can be rewritten as \ben \ch_\lambda(a)\ =\ \
\sum_{w\in W}\, \hd_w\, \left(\, a^\lambda \,\right)\ ,
\label{chDemd}\een or \ben \hch\ =\ \sum_{w\in W}\, \hd_w\ .
\label{sumdw}\een

We will now apply the Demazure character formulas to the
calculation of character generators, following \cite{Ga}. In his
remarkable paper, Gaskell discovered some of the Demazure results
on characters independently, and applied them to character
generators. The motivation was to find formulas that did not
involve negative terms and cancellations, such as the general one
(\ref{PSf}) due to Patera and Sharp \cite{PS}. The relevant minus
signs can be traced to the $\det w$ factor in the Weyl character
formula (\ref{wcf}). As illustrated by the $A_2\cong su(3)$
example above, however, the Demazure character formula can avoid
such negative terms, and so can lead to more useful formulas for
$\rX$.

To save writing, let us introduce the notation \ben \lfloor x
\rfloor\ :=\ (1-x)^{-1}\ =\ \sum_{n=0}^\infty x^n\ .
\label{floor}\een The generating function for highest weights can
be written as \ben {\rm H}(L,a)\ :=\ \prod_{\Lambda\in
F}\,(1-L^\Lambda a^\Lambda)^{-1}\ =\ \prod_{\Lambda\in F}\,
\lfloor L^\Lambda a^\Lambda \rfloor\ , \label{Hdef}\een and the
generating function of interest is then \ben \rX\ =\ \hch \left(\
\prod_{\Lambda\in F}\,(1-L^\Lambda a^\Lambda)^{-1}\ \right)\ =\
\hch \left(\, {\rm H} \,\right)\ = \hD_L\left(\, {\rm H}
\,\right)\ . \label{Gi}\een This general formula for the character
generator was first written by Gaskell \cite{Ga}. Choosing a
reduced decomposition of $\hD_L$, $\rX$ can be calculated by
successive applications of the basic Demazure operators $\hD_i$.

To proceed, Gaskell \cite{Ga} derived the product rule \ben
\hD_i\,\left(F\,G \right)\ =\ (\hD_i F)\,G\ +\ (\hr_iF)\,(\hd_iG)\
. \label{Dprodi}\een This also implies \ben \hD_i\,\left(F\,G
\right)\ =\ F\,(\hD_i G)\ +\ (\hd_iF)\,(\hr_iG)\ .
\label{Dprodii}\een In terms of the operators $\hd_i$, these
identities read as \ben \hd_i\,\left(F\,G \right)\ =\ (\hd_i
F)\,G\ +\ (\hr_iF)\,(\hd_iG)\ \label{dprodi}\een and \ben
\hd_i\,\left(F\,G \right)\ =\ F\,(\hd_i G)\ +\ (\hd_iF)\,(\hr_iG)\
.\label{dprodii}\een For us, the most useful of these product
rules will be (\ref{Dprodii}).

\begin{figure}[t]
\begin{minipage}{10cm}
\vskip-1cm
\begin{center}
\epsfxsize=15cm \epsfbox{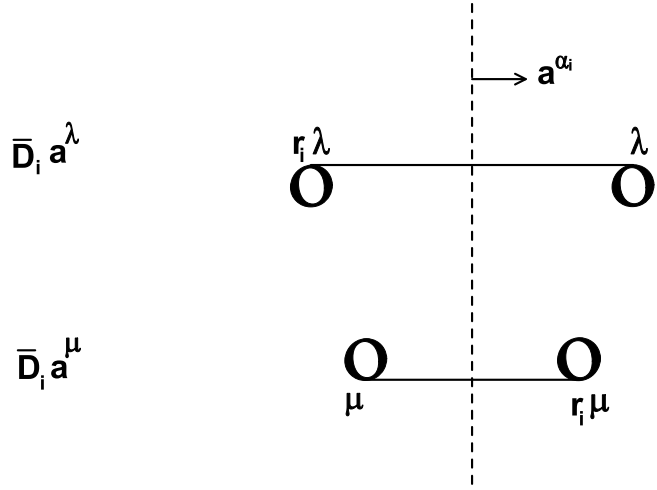} \vskip-13.5cm \caption{\quad The
action of the operator $\Db_i = \hD_i\, a^{-\alpha_i}$.}
\label{DBbar} 
\end{center}
\end{minipage}
\end{figure}

To apply the Demazure operators on individual factors of
$\rH(L,a)$ and the results, we need \ben \hd_i\,\lf F \rf\ =\ \lf
F \rf\, \hd_iF \lf \hr_i F \rf\, \ \label{disF}\een or \ben
\hD_i\,\lf F \rf\ =\ \lf F\rf\ +\ \lf F \rf\, \hd_iF \lf \hr_i F
\rf\, \ . \label{DisF}\een $\hD_i$ always acts to produce an
$r_i$-invariant expression. The right-hand-side of the last result
is therefore $r_i$-invariant, although it is not obvious. A
manifestly invariant formula can be written, however, as \ben
\hD_i\,\lf F \rf\ =\ \lf F\rf\, \left(\, 1 +  \Db_i F
\,\right)\,\lf \hr_i F \rf\, \ . \label{DisFdmr}\een Since we will
need to write it often, we have defined \ben \Db_i\ :=\ \hd_i -
\hr_i\ . \label{Dbdef}\een From the expression (\ref{Dia}), we can
show that \ben  \Db_i\ =\ \hD_i\, a^{-\alpha_i}\ ,
\label{DbhD}\een demonstrating that $\Db_i$, too, generates
$r_i$-invariant expressions. The action of $\Db_i$ is depicted in
Fig. \ref{DBbar}.

Before treating examples, let us use Demazure operators to
re-derive the new formula $\rX=\rY/\rZ$, with numerator and
denominator given by (\ref{Yiii}) and (\ref{XYZ}), respectively.
Re-write (\ref{Gi}) as \ben \rX\ =\ \hD_L\left(\, \frac{{\cal
Y}H}{\cal Y} \,\right)\ =\ \hD_L\left(\, {\rZ}^{-1}\,{\cal Y}
\,\right)\ .\label{XYHZ}\een Since $\rZ$ is $W$-invariant, it is
annihilated by all the $\hd_i$. The product rule (\ref{Dprodii})
therefore yields \ben \rX\ =\ \rZ^{-1}\, \hD_L\, ({\cal Y})\ =\
\rZ^{-1}\, \hch\, ({\cal Y})\ ,\label{XZchY}\een the desired
result.


\vskip1cm
\subsection{Rank-Two Simple Lie Algebras}

The calculations quickly become unwieldy with increasing rank. For
simplicity, we will restrict to consideration of the simple Lie
algebras of rank two. We believe our results are indicative of
general properties of the character generators of the simple Lie
algebras, however.

We should point out that we make certain choices when we perform
our Demazure calculations, such as the order of factors in the
highest-weight generating function $\rH$, the reduced
decomposition of $w_L$, and which of the product rules
(\ref{Dprodi}-\ref{dprodii}) we apply. Of course, none of these
choices changes the final result, but they can simplify the
calculations substantially, and affect the way the final answer is
expressed. It will become clear in Sect. 5 why we make the choices
we do, when a connection with graphs is established.

\subsubsection{$A_2$}

Choosing the reduced decomposition $w_L=r_2r_1r_2$, we need to
calculate \bea \rX\ =& \hD_L\,{\rm H}=\ \hD_2\hD_1\hD_2\, (\lf
L_2a_2\rf\, \lf L_1a_1\rf\ )  .\qq\qq\qq\label{AiiXDi}\eea First
consider the application of $\hD_2$. Since $\lf L_1a_1\rf$ is
$\hr_2$-invariant, it is unaffected. So, \bea \hD_2\, (\lf
L_2a_2\rf\, \lf L_1a_1\rf) = \lf L_2a_2\rf(1+\Db_2L_2a_2) \lf
L_2a_1a_2^{-1}\rf\, \lf L_1a_1\rf\q \nn =\ \lf L_2a_2\rf\, \lf
L_2a_1a_2^{-1}\rf\, \lf L_1a_1\rf\ .\qq\qq\label{AiiXDii}\eea
where we used (\ref{DisFdmr}), and  $\Db_2L_2a_2=0$.

When we apply $\hD_1$, $\lf L_2a_2 \rf$ is untouched: $\hD_1\lf
L_2a_2 \rf = \lf L_2a_2 \rf.$ Using the product rule
(\ref{Dprodii}), we get \bea \hD_1\Big(\hD_2\, (\lf L_2a_2\rf\,
\lf L_1a_1\rf)\Big) =\ \underline{\lf L_2a_2\rf}\, \lf
L_2a_1a_2^{-1}\rf\,\lf L_1a_1\rf(1+\Db_1L_1a_1)\lf
L_1a_1^{-1}a_2\rf\nn +\ \underline{\lf L_2a_2\rf}\, \lf
L_2a_1a_2^{-1}\rf \hd_1L_2a_1a_2^{-1} \lf L_2a_1^{-1}\rf\, \lf
L_1a_1^{-1}a_2\rf\ ,\qq \label{AiiXDiii}\eea which simplifies
since $\Db_1L_1a_1=0$. At this point, we can save effort by
anticipating the application of $\hD_2$, and rewriting the result
as \bea \hD_1\Big(\hD_2\, (\lf L_2a_2\rf\, \lf L_1a_1\rf)\Big) =\
\underline{\hD_2\,( \lf L_2a_2\rf\, \lf L_1a_1\rf)}\,\, \lf
L_1a_1^{-1}a_2\rf\qq\qq\nn +\ \underline{\lf L_2a_2\rf\, \lf
L_2a_1a_2^{-1}\rf \, L_2a_1^{-1}\lf L_2a_1^{-1}\rf}\, \lf
L_1a_1^{-1}a_2\rf\ ,\qq\label{AiiXDiv}\eea since
$\hd_1L_2a_1a_2^{-1}=L_2a_1^{-1}$.  In this last expression, terms
that are $\hr_2$-invariant are made plain by underlines, and all
such terms are annihilated by $\hD_2$. This procedure will save
considerable work in more complicated cases.

For $A_2$, we therefore find \bea \hD_2\bigg(\hD_1\Big(\hD_2\,
(\lf L_2a_2\rf\,\lf L_1a_1\rf)\Big)\bigg)\QQ\QQ\q\nn =\
\underline{\hD_2\, (\lf L_2a_2\rf\, \lf L_1a_1\rf)}\lf
L_1a_1^{-1}a_2\rf(1+\Db_2L_1a_1^{-1}a_2)\lf L_1a_2^{-1}\rf\q\nn +\
\underline{\lf L_2a_2\rf\, \lf L_2a_1a_2^{-1}\rf\,L_2a_1^{-1} \lf
L_2a_1^{-1}\rf}\,\lf L_1a_1^{-1}a_2\rf\qq\nn
\QQ\times(1+\Db_2L_1a_1^{-1}a_2)\lf L_1a_2^{-1}\rf
.\qq\label{AiiXDv}\eea Since  $\Db_2L_1a_1^{-1}a_2$  vanishes, the
final result is \bea \rX\ =\ \hD_2\hD_1\hD_2\, (\lf L_2a_2\rf\,
\lf L_1a_1\rf)\QQ\QQ\qq\nn =\ \lf L_2a_2\rf\, \lf
L_2a_1a_2^{-1}\rf\, \lf L_1a_1\rf\,\lf L_1a_1^{-1}a_2\rf\, \lf
L_1a_2^{-1}\rf\q\QQ\nn +\ \lf L_2a_2\rf\, \lf L_2a_1a_2^{-1}\rf
L_2a_1^{-1} \lf L_2a_1^{-1}\rf\, \lf L_1a_1^{-1}a_2\rf\, \lf
L_1a_2^{-1}\rf\ .\qq\label{AiiXDvi}\eea

Incidentally, the same result can be found by \bea \rY\ =\ \hD_L\,
{\cal Y}\ =\ \hD_1\hD_2\hD_1\,{\cal Y} \ \ \ \ \ \qq\qq\qq\qq\nn \
=\ \hD_1\hD_2\big\{ (1-L_2a_1a_2^{-1})(1-L_2a_1^{-1}) \big[
\hD_1(1-L_1a_1^{-1}a_2)\big] (1-L_2a_2^{-1}) \big\} \nn =\
\hD_1\hD_2\big\{ (1-L_2a_1a_2^{-1})(1-L_2a_1^{-1}) (1-L_2a_2^{-1})
\big\} \qq\qq\nn =\ \hD_1\big\{ (1-L_2a_1^{-1})(1-L_1a_2)^{-1} +\
(1-L_2a_2)(1-L_2a_1^{-1})L_1a_2^{-1} \big\} \ \ \nn =\
(1-L_1a_2^{-1})\ +\ (1-L_2a_2) L_1a_2^{-1}\ ,\
\qq\qq\qq\label{AiiYD}\eea for example.

The structure of the generating functions is more easily seen if
we write \bea A=L_2a_2,\ B=L_2a_1a_2^{-1},\ C=L_2a_1^{-1},\nn
D=L_1a_1,\ E=L_1a_1^{-1}a_2,\ F=L_1a_2^{-1}\ , \label{ABCdefA}\eea
so that \ben \rX\ =\ \lf A \rf\, \lf B \rf\,  \Big(\, \lf D \rf\,
+ C\lf C \rf\, \Big)\, \lf E \rf\, \lf F \rf\ . \label{ABCXA}\een

First, reconsider the 3 choices of incompatible product displayed
in (\ref{fpAii}). They are $DC$, $EB$ and $FA$, respectively.
Expanding the expression of (\ref{ABCXA}) using (\ref{floor})
results in no terms involving the product $DC$. Choosing $DC$ as
the sole incompatible product for the $A_2$ case therefore leads
to (\ref{ABCXA}). Rewriting that expression as $Z^{-1}\, (1-DC)$,
where \ben Z^{-1}\ =\ \lf A \rf\, \lf B \rf\, \lf C \rf\, \lf D
\rf\, \lf E \rf\, \lf F \rf\, \ ,\label{AiiZ}\een makes clear that
the incompatibility $DC$ is related to (\ref{ABCXA}). Similarly,
we can write \bea  \rX\ =\  \lf C \rf\, \lf A \rf\, \Big(\, \lf E
\rf\, + B\lf B \rf\, \Big)\, \lf F \rf\, \lf D \rf\ =\ Z^{-1}\,
(\, 1 - EB \,)\quad\nn\ =\  \lf B \rf\, \lf C \rf\, \Big(\, \lf F
\rf\, + A\lf A \rf\, \Big)\, \lf D \rf\, \lf E \rf\ =\ Z^{-1}\,
(\, 1 - FA \,)\ , \label{fprodAii}\eea corresponding to the other
2 choices $EB$ and $FA$, respectively, for the incompatible
product.

\subsubsection{$B_2$}

The longest element of the $B_2$ Weyl group has the reduced
decompositions $w_L= r_1r_2r_1r_2 = r_2r_1r_2r_1$. Choosing the
first, we write \bea \rX\ =& \hD_L\,{\rm H}=\
\hD_1\hD_2\hD_1\hD_2\, (\lf L_2a_2\rf\, \lf L_1a_1\rf)\
.\qq\qq\label{BiiXDi}\eea Applying (\ref{DisFdmr}), we get\bea
\hD_2\, (\lf L_2a_2\rf\, \lf L_1a_1\rf) =& \lf L_2a_2\rf\, \lf
L_2a_1a_2^{-1}\rf\, \lf L_1a_1\rf\ ,\qq\q\label{BiiXDii}\eea so
that \bea \hD_1\hD_2\, (\lf L_2a_2\rf\, \lf L_1a_1\rf) \ =\
\QQ\QQ\nn\lf L_2a_2\rf\, \lf L_2a_1a_2^{-1}\rf\, \lf L_1a_1\rf\,
(1+\Db_1L_1a_1)\, \lf L_1a_1^{-1}a_2^{2}\rf\qq\nn +\ \lf
L_2a_2\rf\, \lf L_2a_1a_2^{-1}\rf\hd_1L_2a_1a_2^{-1} \lf
L_2a_1^{-1}a_2\rf\, \lf L_1a_1^{-1}a_2^{2}\rf\
,\q\label{BiiXDiii}\eea where (\ref{Dprodii}) was used. This
simplifies because $\Db_1L_1a_1=0$. Here again we can rewrite for
manifest $r_2$-invariance, before applying $\hD_2$: \bea
\hD_1\hD_2\, (\lf L_2a_2\rf\,\lf L_1a_1\rf)\ =\ \underline{\hD_2\,
(\lf L_2a_2\rf\, \lf L_1a_1\rf)}\, \lf
L_1a_1^{-1}a_2^{2}\rf\QQ\nn\qq +\ \underline{\lf L_2a_2\rf\, \lf
L_2a_1a_2^{-1}\rf}\, L_2a_1^{-1}a_2 \lf L_2a_1^{-1}a_2\rf\, \lf
L_1a_1^{-1}a_2^{2}\rf\ ,\qq\label{BiiXDiv}\eea where we have
substituted $\hd_1L_2a_1a_2^{-1} = L_2a_1^{-1}a_2$. Then using
$\hd_2\hD_2=0$, and that $\hd_2$ annihilates all $r_2$-invariant
terms, we get \bea \hD_2\hD_1\hD_2\, (\lf L_2a_2\rf\,\lf
L_1a_1\rf)\QQ\QQ\qq\nn =\ \underline{\hD_2\, (\lf L_2a_2\rf\, \lf
L_1a_1\rf)}\, \lf
L_1a_1^{-1}a_2^{2}\rf(1+\Db_2L_1a_1^{-1}a_2^{2})\lf
L_1a_1a_2^{-2}\rf\qq\nn \ +\ \underline{\, \lf L_2a_2\rf\, \lf
L_2a_1a_2^{-1}\rf }\,\, L_2a_1^{-1}a_2 \lf L_2a_1^{-1}a_2\rf\, \lf
L_1a_1^{-1}a_2^{2}\rf\qq\qq\nn \times
(1+\Db_2L_1a_1^{-1}a_2^{2})\lf L_1a_1a_2^{-2}\rf\qq\q\nn\ +\
\underline{\, \lf L_2a_2\rf\, \lf L_2a_1a_2^{-1}\rf }\,\,
L_2a_1^{-1}a_2 \lf L_2a_1^{-1}a_2\rf \hd_2L_2a_1^{-1}a_2 \qq\q\nn
\times\lf L_2a_2^{-1}\rf\, \lf L_1a_1a_2^{-2}\rf\qq\q\nn\ +\
\underline{\, \lf L_2a_2\rf\, \lf L_2a_1a_2^{-1}\rf \,}\,
\hd_2L_2a_1^{-1}a_2 \lf L_2a_2^{-1}\rf\, \lf L_1a_1a_2^{-2}\rf \
.\qq\qq \label{BiiXDv}\eea Using (\ref{BiiXDiv}), this can be
rewritten as \bea \underline{\hD_1\hD_2\, (\lf L_2a_2\rf\,\lf
L_1a_1\rf)(1+\Db_2L_1a_1^{-1}a_2^{2})}\lf
L_1a_1a_2^{-2}\rf\qq\q\nn +\ \ \underline{\, \lf L_2a_2\rf\, \lf
L_2a_1a_2^{-1}\rf\, \lf L_2a_1^{-1}a_2\rf\, \hd_2L_2a_1^{-1}a_2\,
\lf L_2a_2^{-1}\rf\,}\, \lf L_1a_1a_2^{-2}\rf\
.\label{BiiXDvi}\eea Notice that $\Db_2L_1a_1^{-1}a_2^{2} = L_1$.
Taking account of the $r_1$-invariant terms, it is now simple to
derive \bea \hD_1\hD_2\hD_1\hD_2\, (\lf L_2a_2\rf\, \lf
L_1a_1\rf)\QQ\QQ\qq\nn =\underline{\hD_1\hD_2\, (\lf L_2a_2\rf\,
\lf L_1a_1\rf)(1+\Db_2L_1a_1^{-1}a_2^{2})}\lf
L_1a_1a_2^{-2}\rf\,\lf L_1a_1^{-1}\rf\q\nn +\ \underline{\lf
L_2a_2\rf\, \lf L_2a_1a_2^{-1}\rf\,\lf L_2a_1^{-1}a_2\rf
\hd_2L_2a_1^{-1}a_2\lf L_2a_2^{-1}\rf}\,\qq\q\nn\times \lf
L_1a_1a_2^{-2}\rf\,\lf L_1a_1^{-1}\rf\ .\qq\label{BiiXDvii}\eea

Finally, substituting for $\hD_1\hD_2\, (\lf L_2a_2\rf\, \lf
L_1a_1\rf)$ from above, we get \bea \hD_1\hD_2\hD_1\hD_2\, (\lf
L_2a_2\rf\, \lf L_1a_1\rf)\QQ\QQ\qq\nn =\ \lf L_2a_2\rf\,\lf
L_2a_1a_2^{-1}\rf\, \lf L_1a_1\rf\, \lf
L_1a_1^{-1}a_2^{2}\rf\qq\qq\qq\q\nn\times(1+\Db_2L_1a_1^{-1}a_2^{2})\lf
L_1a_1a_2^{-2}\rf\,\lf L_1a_1^{-1}\rf\nn +\ \lf L_2a_2\rf\, \lf
L_2a_1a_2^{-1}\rf L_2a_1^{-1}a_2 \lf L_2a_1^{-1}a_2\rf\, \lf
L_1a_1^{-1}a_2^{2}\rf\qq\qq\nn\times(1+\Db_2L_1a_1^{-1}a_2^{2})\lf
L_1a_1a_2^{-2}\rf\,\lf L_1a_1^{-1}\rf\nn +\ \lf L_2a_2\rf\, \lf
L_2a_1a_2^{-1}\rf\,\lf L_2a_1^{-1}a_2\rf L_2a_2^{-1}\lf
L_2a_2^{-1}\rf\,\qq\q\nn\times \lf L_1a_1a_2^{-2}\rf\,\lf
L_1a_1^{-1}\rf\ .\q\label{BiiXDviii}\eea

Using the notation \bea A=L_2a_2,\ B=L_2a_1a_2^{-1},\
C=L_2a_1^{-1}a_2, D=L_2a_2^{-1},\ \  \nn  E=L_1a_1,\
F=L_1a_1^{-1}a_2^2,\ G=L_1 a_1a_2^{-2} ,\ H=L_1a_1^{-1}  \ ,
\label{ABCdefB}\eea the $B_2$ character generator takes a compact
form. Defining \ben \lf\, A\,B \,\rf\ :=\ \lf A \rf\, \lf B \rf\,
\ ,\label{bABb}\een and similarly for more than two factors, we
can write  \bea \rX\ =\ \lf\, A \, B \, C\, \rf\, D\,\lf\, D\, G\,
H\, \rf\QQ\q\nn \ +\ \lf\, A\, B\, \rf\, C\, \lf\, C\, F\,\rf\,
(1+z)\, \lf\, G\, H\, \rf\qq \nn \ +\ \lf\, A\,  B\,  E\, F\,
\rf\, (1+z)\, \lf\, G\, H\, \rf\ .\q\label{ABCXB}\eea Here we have
also defined $z\ :=\ \Db_2L_1a_1^{-1}a_2^{2}\ =\ L_1$ for the sole
inside generator required for the $B_2$ generating function.

\begin{figure}[t]
\begin{minipage}{10cm}
\vskip-2cm
\begin{center}
\epsfxsize=17cm \epsfbox{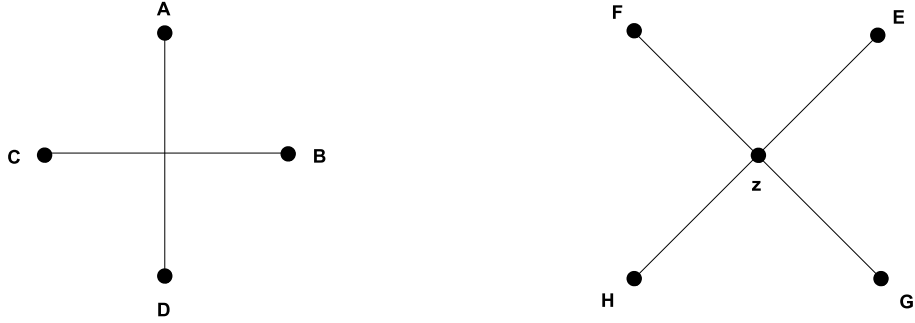} \vskip-16.5cm
\caption{\quad The $B_2$ fundamental weight diagrams. The weights
are labelled by the corresponding elements of the integrity basis
$I_X$.}
\label{B2fund} 
\end{center}
\end{minipage}
\end{figure}

The weights of the generators $A-H$ and $z$ are depicted in Fig.
\ref{B2fund}. For all character generators, the generator weights
fill out the $r$ fundamental weight diagrams of the relevant
rank-$r$ simple Lie algebra. The two fundamental weight diagrams
of $B_2$ are shown in the Figure.

The form of the character generator can be related to the set of
incompatible products obtained above. In terms of the integrity
basis elements, the choice (\ref{fpBii}) gives $\{EC, ED, FD, zD
\}$ as the set of incompatible products. It is easily seen that
the expression (\ref{ABCXB}) does not contain these products, but
does contain all other products quadratic in the elements of
$\{A,B,\ldots,F,z\}$.

\subsubsection{$G_2$}

For $G_2$, $w_L=r_1r_2r_1r_2r_1r_2 = r_2r_1r_2r_1r_2r_1$ are the
two reduced decompositions of the longest element of the Weyl
group.  We will calculate \bea \rX\ = \hD_L\,{\rm
H}=\hD_1\hD_2\hD_1\hD_2\hD_1\hD_2\, (\lf L_2a_2\rf\, \lf
L_1a_1\rf) .\qq\qq\label{GiiXD}\eea

First, since $\Db_2L_2a_2 = 0$, \bea \hD_2\, (\lf L_2a_2\rf\, \lf
L_1a_1\rf) =\lf L_2a_2\rf\, \lf L_2a_1a_2^{-1}\rf\, \lf L_1a_1\rf\
,\qq\q\label{GiiXDi}\eea using (\ref{DisFdmr}). Since $\lf L_2a_2
\rf$ is $r_1$-invariant, applying $\hD_1$ to this last result
gives \bea \hD_1\hD_2\, (\lf L_2a_2\rf\, \lf L_1a_1\rf)\QQ\QQ\q\nn
=\ \lf L_2a_2\rf\, \lf L_2a_1a_2^{-1}\rf\, \lf L_1a_1\rf\, \lf
L_1a_1^{-1}a_2^{3}\rf\qq\qq \nn +\ \lf L_2a_2\rf\, \lf
L_2a_1a_2^{-1}\rf\hd_1L_2a_1a_2^{-1} \lf L_2a_1^{-1}a_2^{2}\rf\,
\lf L_1a_1^{-1}a_2^{3}\rf\ ,\q\label{GiiXDii}\eea using the
product rule (\ref{Dprodii}), as usual. $\hd_1L_2a_1a_2^{-1}$ is
just $L_2a_1^{-1}a_2^{2}$.

We will try to keep track of the invariant terms as we proceed.
This will simplify the calculations greatly. We rewrite the above
equation as \bea \hD_1\hD_2\, (\lf L_2a_2\rf\, \lf
L_1a_1\rf)\QQ\QQ\qq\nn\ =\ \underline{\hD_2\, (\lf L_2a_2\rf\, \lf
L_1a_1\rf) }\,\lf L_1a_1^{-1}a_2^{3}\rf\,\qq\qq\qq\nn +\
\underline{\lf L_2a_2\rf\,\lf L_2a_1a_2^{-1}\rf} \
L_2a_1^{-1}a_2^{2} \lf L_2a_1^{-1}a_2^{2}\rf\,\lf
L_1a_1^{-1}a_2^{3}\rf\ ,\qq\label{GiiXDiii}\eea where the
underlines indicate $r_2$-invariant factors. Applying $\hD_2$ then
gives  \bea \hD_2\hD_1\hD_2\, (\lf L_2a_2\rf\, \lf
L_1a_1\rf)\QQ\QQ\qq\nn =\ \underline{ \hD_2\, (\lf L_2a_2\rf\, \lf
L_1a_1\rf)\ }\,\lf L_1a_1^{-1}a_2^{3}\rf\,(1+\Db_2
L_1a_1^{-1}a_2^{3}) \lf L_1a_1^{2}a_2^{-3}\rf\q\nn +\
\underline{\lf L_2a_2\rf\,\lf L_2a_1a_2^{-1}\rf}\ \,
L_2a_1^{-1}a_2^{2} \lf L_2a_1^{-1}a_2^{2} \rf\,\lf
L_1a_1^{-1}a_2^{3}\rf\,\qq\q\nn\times(1+\Db_2 L_1a_1^{-1}a_2^{3})
\lf L_1a_1^{2}a_2^{-3}\rf \nn +\ \underline{\lf L_2a_2\rf\, \lf
L_2a_1a_2^{-1}\rf} \, \lf L_2a_1^{-1}a_2^{2}\rf\,\hd_2
L_2a_1^{-1}a_2^{2}\, \lf L_2a_1a_2^{-2}\rf\,\lf
L_1a_1^{2}a_2^{-3}\rf\ .\label{GiiXDiv}\eea Using
(\ref{GiiXDiii}), this expression can be simplified to
\bea\hD_2\hD_1\hD_2\, (\lf L_2a_2\rf\, \lf L_1a_1\rf)\QQ\QQ\qq\nn
=\ \underline{\hD_1\hD_2\, (\lf L_2a_2\rf\, \lf L_1a_1\rf)}
(1+\Db_2 L_1a_1^{-1}a_2^{3}) \lf L_1a_1^{2}a_2^{-3}\rf\qq\q\nn +\
\underline{\lf L_2a_2\rf\,\lf L_2a_1a_2^{-1}\rf\,\lf
L_2a_1^{-1}a_2^{2}\rf} \, \hd_2L_2a_1^{-1}a_2^{2}\lf
L_2a_1a_2^{-2}\rf\, \lf L_1a_1^{2}a_2^{-3}\rf\ ,\label{GiiXDv}\eea
where the underlines now indicate $r_1$-invariant factors, in
preparation for the application of $\hD_1$.

When $\hD_1$ is applied, we get \bea\hD_1\hD_2\hD_1\hD_2\, (\lf
L_2a_2\rf\, \lf L_1a_1\rf)\QQ\QQ\qq\nn =\
\underline{\hD_1\hD_2\,(\lf L_2a_2\rf\, \lf L_1a_1\rf)} (1+\Db_2
L_1a_1^{-1}a_2^{3})\hD_1\lf L_1a_1^{2}a_2^{-3}\rf\qq\q\nn +\
\underline{\hD_1\hD_2\,(\lf L_2a_2\rf\,\lf L_1a_1\rf)}
\hd_1\Db_2(L_1a_1^{-1}a_2^{3})\lf L_1a_1^{-2}a_2^{3}\rf\qq\qq\nn
+\ \ \underline{\lf L_2a_2\rf\,\lf L_2a_1a_2^{-1}\rf\,\lf
L_2a_1^{-1}a_2^{2}\rf} \hd_2 L_2a_1^{-1}a_2^{2}\lf
L_2a_1a_2^{-2}\rf\,\qq\nn\times\hD_1\lf L_1a_1^{2}a_2^{-3}\rf\q\nn
+\ \underline{\lf L_2a_2\rf\,\lf L_2a_1a_2^{-1}\rf\, \lf
L_2a_1^{-1}a_2^{2}\rf} \hd_2 L_2a_1^{-1}a_2^{2} \lf
L_2a_1a_2^{-2}\rf\qq\nn\times\hd_1L_2a_1a_2^{-2}\lf
L_2a_1^{-1}a_2\rf\, \lf L_1a_1^{-2}a_2^{3}\rf\q\nn +\
\underline{\lf L_2a_2\rf\, \lf L_2a_1a_2^{-1}\rf\,\lf
L_2a_1^{-1}a_2^{2}\rf} \hd_1\hd_2 L_2a_1^{-1}a_2^{2}\qq\q\nn\times
\lf L_2a_1^{-1}a_2\rf\, \lf L_1a_1^{-2}a_2^{3}\rf\
.\qq\label{GiiXDvi}\eea Using the results above, this becomes
\bea\hD_1\hD_2\hD_1\hD_2\,(\lf L_2a_2\rf\,\lf
L_1a_1\rf)\QQ\QQ\qq\nn =\ \underline{ \hD_2\hD_1\hD_2\,(\lf
L_2a_2\rf\,\lf L_1a_1\rf )} (1+\Db_1L_1a_1^{2}a_2^{-3})\lf
L_1a_1^{-2}a_2^{3}\rf \qq\q\nn +\ \Big(\hD_1\hD_2\,(\lf
L_2a_2\rf\,\lf L_1a_1\rf)\Big) \hd_1\Db_2(L_1a_1^{-1}a_2^{3})  \lf
L_1a_1^{-2}a_2^{3}\rf\qq\qq\nn +\ \underline{\lf L_2a_2\rf\, \lf
L_2a_1a_2^{-1}\rf\, \lf
L_2a_1^{-1}a_2^{2}\rf(1+\hD_2L_2a_1^{-1}a_2^{2})\lf
L_2a_1a_2^{-2}\rf} \qq\nn\times L_2a_1^{-1}a_2\lf
L_2a_1^{-1}a_2\rf\,\lf L_1a_1^{-2}a_2^{3}\rf\
,\qq\label{GiiXDvii}\eea where now the underlines indicate
$r_2$-invariant factors.

Letting $\hD_2$ act, we obtain \bea
\hD_2\hD_1\hD_2\hD_1\hD_2\,(\lf L_2a_2\rf\,\lf
L_1a_1\rf)\QQ\QQ\qq\nn =\ \underline{\hD_2\hD_1\hD_2\, (\lf
L_2a_2\rf\,\lf L_1a_1\rf)} (1+\Db_1L_1a_1^{2}a_2^{-3})\hD_2\lf
L_1a_1^{-2}a_2^{3}\rf \qq\nn +\  \underline{\hD_2\hD_1\hD_2\,(\lf
L_2a_2\rf\,\lf L_1a_1\rf)} \hd_2\Db_1(L_1a_1^{2}a_2^{-3})\lf
L_1a_1a_2^{-3}\rf\qq\qq\nn +\ \Big(\hD_1\hD_2\,(\lf L_2a_2\rf\,\lf
L_1a_1\rf)\Big) \hd_1\Db_2(L_1a_1^{-1}a_2^{3})\,\hD_2\lf
L_1a_1^{-2}a_2^{3}\rf\,\qq\qq\nn +\ \Big(\hD_1\hD_2\,(\lf
L_2a_2\rf\,\lf L_1a_1\rf)\Big)
\Db_2\hd_1\Db_2(L_1a_1^{-1}a_2^{3})\lf L_1a_1a_2^{-3}\rf\qq\qq\nn
+\ \Big(\hD_2 \hD_1\hD_2\,(\lf L_2a_2\rf\,\lf L_1a_1\rf)
\Big)\hr_2\hd_1\Db_2(L_1a_1^{-1}a_2^{3})\lf
L_1a_1a_2^{-3}\rf\qq\q\nn +\ \underline{\lf L_2a_2\rf\, \lf
L_2a_1a_2^{-1}\rf\, \lf L_2a_1^{-1}a_2^{2}\rf
(1+\Db_2L_2a_1^{-1}a_2^{2})\lf L_2a_1a_2^{-2}\rf} \qq\nn\times
L_2a_1^{-1}a_2\lf L_2a_1^{-1}a_2\rf\,\hD_2\lf
L_1a_1^{-2}a_2^{3}\rf\q\nn +\ \underline{\lf L_2a_2\rf\, \lf
L_2a_1a_2^{-1}\rf\, \lf
L_2a_1^{-1}a_2^{2}\rf(1+\Db_2L_2a_1^{-1}a_2^{2})\lf
L_2a_1a_2^{-2}\rf} \qq\nn\times \lf
L_2a_1^{-1}a_2\rf\,\hd_2L_2a_1^{-1}a_2\lf L_2a_2^{-1}\rf\,\lf
L_1a_1a_2^{-3}\rf\ .\q\label{GiiXDviii}\eea Here we have used our
usual product rule (\ref{Dprodii}), but also the identity \ben F\,
(\hd_iG)\ +\ (\hd_iF)\,(\hr_iG)\ =\ F\, (\Db_iG)\ +\
(\hD_iF)\,(\hr_iG)\ ,\label{dDbid}\een so that the $\big(\hD_2
\hD_1\hD_2\,(\lf L_2a_2\rf\,\lf L_1a_1\rf) \big)$ term results.

This expression can be simplified. For example,
$\hd_2\Db_1(L_1a_1^{2}a_2^{-3})=0$. Completing the calculation,
and referring to (\ref{GiiXDvii}), yields \bea
\hD_2\hD_1\hD_2\hD_1\hD_2\,\lf L_2a_2\rf\,\lf
L_1a_1\rf\QQ\QQ\qq\nn =\ \underline{\hD_1\hD_2\hD_1\hD_2\, \lf
L_2a_2\rf\,\lf L_1a_1\rf} (1+\Db_2L_1a_1^{-2}a_2^{3})\lf
L_1a_1a_2^{-3}\rf\qq\nn +\ \underline{\hD_1\hD_2\, \lf
L_2a_2\rf\,\lf L_1a_1\rf} \Db_2\hd_1\Db_2(L_1a_1^{-1}a_2^{3})\lf
L_1a_1a_2^{-3}\rf\qq\qq\nn +\ \Big(\hD_2\hD_1\hD_2\,(\lf
L_2a_2\rf\,\lf
L_1a_1\rf)\Big)\hr_2\hd_1\Db_2(L_1a_1^{-1}a_2^{3})\lf
L_1a_1a_2^{-3}\rf\qq\q\nn +\ \underline{\lf L_2a_2\rf\,\lf
L_2a_1a_2^{-1}\rf\, \lf
L_2a_1^{-1}a_2^{2}\rf(1+\Db_2L_2a_1^{-1}a_2^{2})\lf
L_2a_1a_2^{-2}\rf}\qq\nn\times \underline{\lf
L_2a_1^{-1}a_2\rf\,L_2a_2^{-1}\lf L_2a_2^{-1}\rf}\lf
L_1a_1a_2^{-3}\rf\ ,\q\label{GiiXDix}\eea where the
$r_1$-invariant terms have been underlined.

Finally, applying $\hD_1$ we obtain \bea
\hD_1\hD_2\hD_1\hD_2\hD_1\hD_2\,(\lf L_2a_2\rf\,\lf
L_1a_1\rf)\QQ\QQ\q\nn =\ \underline{\hD_1\hD_2\hD_1\hD_2\,(\lf
L_2a_2\rf\,\lf L_1a_1\rf)}(1+\Db_2L_1a_1^{-2}a_2^{3})\,\hD_1\lf
L_1a_1a_2^{-3}\rf\ \qq\nn +\ \underline
{\hD_1\hD_2\hD_1\hD_2\,(\lf L_2a_2\rf\,\lf
L_1a_1\rf)}\hd_1\Db_2(L_1a_1^{-2}a_2^{3})\lf L_1a_1^{-1}\rf\qq\nn
+\ \underline{\hD_1\hD_2\,(\lf L_2a_2\rf\,\lf L_1a_1\rf)}
\Db_2\hd_1\Db_2(L_1a_1^{-1}a_2^{3})\,\hD_1\lf L_1a_1a_2^{-3}\rf\
\qq\q\nn +\ \underline{\hD_1\hD_2\,(\lf L_2a_2\rf\,\lf L_1a_1\rf)}
\hd_1\Db_2\hd_1\Db_2(L_1a_1^{-1}a_2^{3})\lf L_1a_1^{-1}\rf\qq\q\nn
+\ \Big(\hD_2\hD_1\hD_2\,(\lf L_2a_2\rf\,\lf
L_1a_1\rf)\Big)\hr_2\hd_1\Db_2(L_1a_1^{-1}a_2^{3})\,\hD_1\lf
L_1a_1a_2^{-3}\rf\,\qq\nn +\ \Big(\hD_2\hD_1\hD_2\,(\lf
L_2a_2\rf\,\lf
L_1a_1\rf)\Big)\Db_1\hr_2\hd_1\Db_2(L_1a_1^{-1}a_2^{3})\lf
L_1a_1^{-1}\rf\qq\nn +\ \Big(\Db_1(\hD_2\hD_1\hD_2\,\lf
L_2a_2\rf\,\lf
L_1a_1\rf)\Big)\hr_1\hr_2\hd_1\Db_2(L_1a_1^{-1}a_2^{3})\lf
L_1a_1^{-1}\rf\qq\nn +\ \underline{\lf L_2a_2\rf\, \lf
L_2a_1a_2^{-1}\rf\, \lf
L_2a_1^{-1}a_2^{2}\rf(1+\hD_2L_2a_1^{-1}a_2^{2})\lf
L_2a_1a_2^{-2}\rf},\q\nn\times \underline{\lf L_1a_1^{-1}a_2\rf
\hd_2L_2a_1^{-1}a_2\lf L_2a_2^{-1}\rf}\hD_1\lf L_1a_1a_2^{-3}\rf\
.\qq\label{GiiXDx}\eea In this expression, both
$\hd_1\Db_2\hd_1\Db_2(L_1a_1^{-1}a_2^{3})$ and
$\Db_1\hr_2\hd_1\Db_2(L_1a_1^{-1}a_2^{3})$ vanish. Also, the
second and seventh summands on the right hand side cancel, because
$\hd_1\Db_2(L_1a_1^{-2}a_2^{3}) = -\hr_1\hr_2\hd_1
\Db_2(L_1a_1^{-1}a_2^{3})$. Therefore, the above result reduces to
\bea \underline{\hD_1\hD_2\hD_1\hD_2\,(\lf L_2a_2\rf\,\lf
L_1a_1\rf)}(1+\Db_2L_1a_1^{-2}a_2^{3})\,\hD_1\lf
L_1a_1a_2^{-3}\rf\, \qq\nn +\ \underline{\hD_1\hD_2\,(\lf
L_2a_2\rf\,\lf L_1a_1\rf)} \Db_2\hd_1\Db_2(L_1a_1^{-1}a_2^{3})
\hD_1\lf L_1a_1a_2^{-3}\rf\, \qq\q\nn +\
\Big(\hD_2\hD_1\hD_2\,(\lf L_2a_2\rf\,\lf
L_1a_1\rf)\Big)\hr_2\hd_1\Db_2(L_1a_1^{-1}a_2^{3})\,\hD_1\lf
L_1a_1a_2^{-3}\rf\,\qq\nn +\ \underline{\lf L_2a_2\rf\, \lf
L_2a_1a_2^{-1}\rf\, \lf
L_2a_1^{-1}a_2^{2}\rf(1+\hD_2L_2a_1^{-1}a_2^{2})\lf
L_2a_1a_2^{-2}\rf}\,\q\nn\times \underline{\lf L_1a_1^{-1}a_2\rf\,
L_2a_2^{-1}\lf L_2a_2^{-1}\rf} \hD_1\lf L_1a_1a_2^{-3}\rf\ ,
\qq\label{GiiXDxi}\eea where $\hD_1\lf L_1a_1a_2^{-3}\rf = \lf
L_1a_1a_2^{-3}\rf\,\lf L_1a_1^{-1}\rf $.

Substituting for $\hD_1\hD_2\hD_1\hD_2\,(\lf L_2a_2\rf\,\lf
L_1a_1\rf)$, $\hD_1\lf L_1a_1a_2^{-3}\rf$, $\hD_1\hD_2\,(\lf
L_2a_2\rf\,\lf L_1a_1\rf)$, and $\hD_2\hD_1\hD_2\,(\lf
L_2a_2\rf\,\lf L_1a_1\rf)$ from above, and defining \bea
A=L_2a_2,\ B=L_2a_1a_2^{-1},\ C=L_2a_1^{-1}a_2^2,\nn
D=L_2a_1a_2^{-2},\ E=L_2a_1^{-1}a_2,\ F= L_2a_2^{-1},\nn
G=L_1a_1,\ H=L_1a_1^{-1}a_2^3,\ I=L_1a_1^{2}a_2^{-3},\nn
J=L_1a_1^{-2}a_2^{3},\ K=L_1a_1a_2^{-3},\ L=L_1a_1^{-1}\ ,
\label{ABCdefG}\eea we can write the character generator for $G_2$
as \bea X\ =\ \hD_1\hD_2\hD_1\hD_2\hD_1\hD_2\,\left(\, \lf
A\rf\,\lf G\rf\,\right)\QQ\QQ\qq\nn =\ \lf  A\rf\,\lf  B\rf\,\lf
G\rf\,\lf H\rf (1+\Db_2  H)\lf  I\rf (1+\Db_1  I)\lf  J\rf
(1+\Db_2  J)\lf K\rf\, \lf  L\rf\qq\nn +\ \ \lf A\rf\,\lf   B\rf\,
C \lf C\rf\,\lf H\rf (1+\Db_2  H)\lf I\rf (1+\Db_1  I)\lf  J\rf
(1+\Db_2 J)\lf K\rf\,\lf L\rf\q\nn +\ \lf A\rf\,\lf   B\rf\,\lf
C\rf\, \hd_2C \lf D\rf\,\lf I\rf (1+\Db_1 I)\lf  J\rf(1+\Db_2
J)\lf K\rf\,\lf L\rf\qq\q\nn +\ \lf A\rf\,\lf B\rf\,\lf  G\rf\,\lf
H\rf \hd_1\Db_2 H \lf J\rf(1+\Db_2  J)\lf  K\rf\,\lf
L\rf\qq\qq\q\nn +\ \lf A\rf\,\lf B\rf\, C \lf  C\rf\,\lf H\rf\,
\hd_1\Db_2 H \lf J\rf (1+\Db_2 J)\lf K\rf\,\lf L\rf\qq\qq\nn +\
\lf  A\rf\,\lf B\rf\,\lf C\rf\,(1 + \Db_2C )\lf D\rf E \lf
E\rf\,\lf J\rf\,(1+\Db_2  J)\lf K\rf\,\lf L\rf\qq\nn +\ \lf
A\rf\,\lf B\rf\,\lf  G\rf\,\lf H\rf\Db_2 \hd_1\Db_2  H\lf
K\rf\,\lf L\rf\qq\qq\qq\q\nn +\ \lf A\rf\,\lf B\rf\, C \lf
C\rf\,\lf H\rf\Db_2\hd_1\Db_2  H\lf K\rf\,\lf L\rf\qq\qq\qq\nn +\
\lf A\rf\,\lf  B\rf\,\lf  G\rf\,\lf H\rf (1+\Db_2  H)\lf
I\rf\hr_2\hd_1\Db_2  H\lf  K\rf\,\lf L\rf\qq\qq\nn +\ \lf
A\rf\,\lf  B\rf\, C \lf  C\rf\,\lf H\rf (1+\Db_2  H)\lf I\rf
\hr_2\hd_1\Db_2  H\lf  K\rf\,\lf L\rf\qq\q\nn +\ \lf A\rf\,\lf
B\rf\,\lf  C\rf\, \hd_2C \lf  D\rf \lf I\rf\hr_2\hd_1\Db_2 H\lf
K\rf\,\lf  L\rf\qq\qq\nn +\ \lf A\rf\,\lf  B\rf\,\lf
C\rf\,(1+\Db_2  C)\lf  D\rf\lf  E\rf  F \lf F\rf\lf K\rf\,\lf
L\rf\ .\qq\label{AGiiXDpf}\eea

\begin{figure}[t]
\begin{minipage}{10cm}
\vskip-1cm
\begin{center}
\epsfxsize=17cm \epsfbox{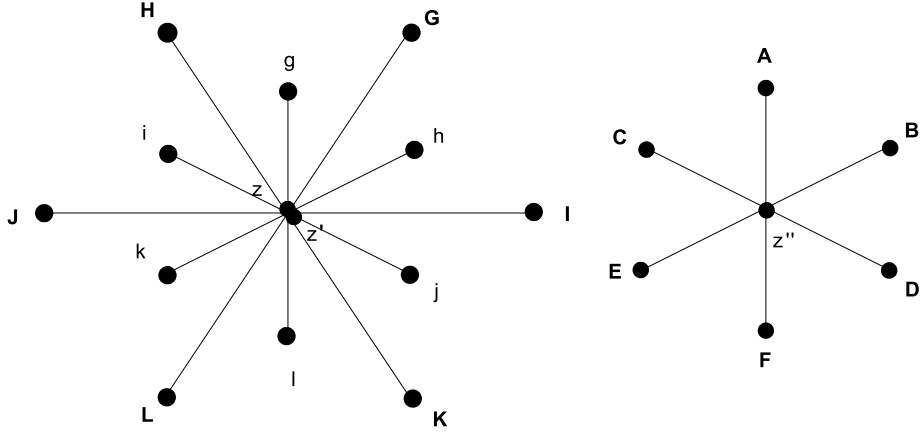} \vskip-15.5cm
\caption{\quad The $G_2$ fundamental weight diagrams. The weights
are indicated by the elements of $I_X$.}
\label{G2fund111} 
\end{center}
\end{minipage}
\end{figure}

The weights of these outside elements (see (\ref{Xbas})) of the
integrity basis are depicted in Fig. \ref{G2fund111}, where the
weight diagrams of the fundamental representations of $G_2$ are
drawn. The other weights are labelled by our notation for the
corresponding inside generators.

The inside generators, the elements of $I_{\rm in}$, make their
appearance when we calculate numerator factors, such as
$\hr_2\hd_1\Db_2  H = j$. The final result is therefore \bea X\ =\
\lf\, A \,  B\,   G\, H\rf \,(1+ \,g+h\, )\, \lf I\rf \,( 1+
\,z'\, )\, \lf J\rf \,( 1+\, k+\ell\, )\, \lf K\,  L\, \rf\qq\nn
+\ \ \lf\, A\, B \,\rf\, C\, \lf C\, H \,\rf \,( 1+ \,g+h\, )\,
\lf I\rf \,( 1+ \,z'\, )\, \lf J\rf \,( 1+\, k+\ell\, )\, \lf\,
K\, L \,\rf\q\nn\qq +\ \lf\, A\, B\, C \,\rf\, (z''+D)\, \lf\, D\,
I \,\rf \,( 1+ \,z'\, )\, \lf J\rf\,( 1+\, k+\ell\, )\, \lf\, K\,
L \,\rf\qq\nn +\ \lf\, A\, B\, G\, H \,\rf \,i\, \lf J\rf\,( 1+\,
k+\ell\, )\,  \lf\, K\, L \,\rf\qq\qq\q\nn +\ \lf\, A\, B\, \rf
\,C\, \lf\, C\, H \,\rf\, \,i\, \lf  J\rf \,( 1+\, k+\ell\, )\,
\lf\, K\, L \,\rf\qq\qq\nn +\ \lf\, A\, B\, C \,\rf \,( 1 +
\,z''\, )\, \lf\, D\,\rf \,E\, \lf\, E\, J \,\rf\, ( 1+\, k+\ell\,
)\,  \lf\, K\, L \,\rf \qq\nn +\ \lf\, A\, B\, G\, H \,\rf \,z\,
\lf\, K\, L \,\rf\ +\ \lf\, A\, B \,\rf\, C\, \lf\, C\, H \,\rf
\,z\, \lf\, K\, L \,\rf\qq\nn +\ \lf\, A\, B\, G\, H \,\rf\, \,(
1+ \,g+h\, ) \,\lf I\rf\, j\, \lf\, K\, L \,\rf \qq\qq\nn +\ \lf\,
A\, B \,\rf\, C\, \lf\, C\, H \,\rf \,(1+ \,g+h\, )\, \lf\, I
\,\rf \, j\, \lf\, K\, L \,\rf \qq\q\nn +\ \lf\, A\, B\, C \,\rf\,
(z''+D)\, \lf D\, I\, \rf\, \,j\, \lf\, K\, L \,\rf \qq\qq\nn +\
\lf\, A\, B\, C \,\rf\,(1+ \,z''\, )\, \lf\, D\, E \,\rf  \,F\,
\lf\, F\, K\, L \,\rf\ .\qq\q\label{AGiiXDf}\eea Here we have
again shortened by using $\lf A\rf \lf B\rf =: \lf AB \rf$, etc.

Now consider the choice of incompatible products underlying this
expression for $\rX$ written in terms of integrity basis elements.
By inspecting (\ref{AGiiXDf}), we can find the incompatibilities
between outer generators: \ben \{ GC, GD, GE, GF, HD, HE, HF, IE,
IF, JF\}\ .\label{incpoo}\een None of these products appears in
the expansion of any of the terms of (\ref{AGiiXDf}).  Therefore,
we write \ben \rY^{(2)}_{\rm out,out}\ =\ G(C+D+E+F) + H(D+E+F) +
I(E+F) + JF\ . \label{GiiYiioo}\een Similarly, for
inner$\cdot$outer incompatible products, we find \bea
\rY^{(2)}_{\rm in,out}\ =\ z''(G+H) + (z+g+h+i)(D+E+F) + z'(E+F)
\nn + j(E+F) + (k+\ell)F + z(I+J) + iI + jJ\
.\qq\label{GiiYiiio}\eea Compatible inner$\cdot$inner products
give \ben \rY^{(2)}_{\rm in,in}\ =\ (z''+g+h)(j+k+\ell+z') +
i(k+\ell+z') + z'(k+\ell) \ . \label{GiiYiiii}\een Substituting
these last 3 results into (\ref{inoutYiiGii}) verifies the result
(\ref{GiiYq}) derived from the general formula (\ref{Yiii}).

\newpage
\section{Character generators, semi-standard tableaux,
posets and graphs}

By (\ref{defch}), the character generator is the generating
function of the multiplicities ${\rm mult}_\lambda(\sigma)$: \ben
\rX\ =\ \sum_{\lambda\in P_\ge}\,\sum_{\sigma\in P}\,
L^\lambda\,a^\sigma\,{\rm mult}_\lambda(\sigma)\
.\label{gfmult}\een Many combinatorial ways of calculating such
multiplicities are known, including those involving Young tableaux
and variants. This ``microscopic'' point of view leads to an
improved understanding of the structure of the character
generators. The first to exploit this fact was Stanley \cite{S},
for the algebras $A_r\cong su(r+1)$. King \cite{K} extended
Stanley's work to include the algebras $C_r\cong sp(2r)$.

Most relevant to us, however, was the connection made explicit by
Baclawski \cite{Ba} to certain partially-ordered sets, or posets,
related to tableaux. A poset ${\cal P}$ is a set, together with a
binary operation (partial order) $\ge$, satisfying reflexivity
($x\ge x, \ \forall x\in {\cal P}$), antisymmetry (if $x\ge y$ and
$y\ge x$, then $x=y$), and transitivity (if $x\ge y$ and $y\ge z$,
then $x\ge z$). It is a partial order because two elements $x,y\in
{\cal P}$ can be incomparable, i.e. neither $x\ge y$ nor $y\ge x$
is true.

The connection with posets was already made in \cite{S}, but much
less directly than in \cite{Ba}. Baclawski emphasized its
importance and wrote explicit formulas in terms of poset objects.
Later these considerations were generalized to all the classical
Lie algebras $A_r,B_r,C_r, D_r$ (or all $su(N),so(N),sp(2N)$) in
\cite{KEi,KEii}, using generalized Young tableaux.\footnote{\ For
a discussion of character generators and tableaux methods with a
different emphasis, see \cite{CCS}.} At about the same time,
Baclawski and Towber \cite{BT} treated the exceptional $G_2$
algebra by introducing a generalization of a poset.

In the remainder of this section, we will treat the algebras
$A_r$, $B_2$ and $G_2$ in turn. This first case is the simplest,
and best understood.

\subsection{$A_r=su(r+1)$}

Certain posets are encoded in the structure of Young tableaux, and
related objects. For example, consider the algebras $A_{r}\cong
su(r+1)$. Their multiplicities ${\rm mult}_\lambda(\sigma)$ equal
the number of semi-standard Young tableaux of shape $\lambda$ and
weight $\sigma$ (see \cite{FH}, e.g.). These Young tableaux can be
constructed by joining together the semi-standard tableaux of the
fundamental representation, and these fundamental tableaux become
the columns of the full semi-standard tableaux. The only
complication is that they must be placed in a certain order.

More precisely, the columns of the semi-standard tableaux, the
fundamental tableaux, are the elements of a poset ${\cal P}$. The
partial order can be encoded in a so-called Hasse diagram, a graph
whose vertices are the elements of the poset, and whose edges
indicate the order (see \cite{St}, e.g.). The poset ${\cal P}$ is
locally finite, meaning it has an order that is completely
determined by its cover relations. $x>y$ is a cover relation if no
poset element $z$ exists such that $x>z>y$.  To every cover
relation $x>y$ of the poset ${\cal P}$, there is an edge $\{x,y\}$
in its Hasse diagram ${\cal H}({\cal P})$.

\begin{figure}[t]
\begin{minipage}{10cm}
\vskip-2cm
\begin{center}
\epsfxsize=17cm \epsfbox{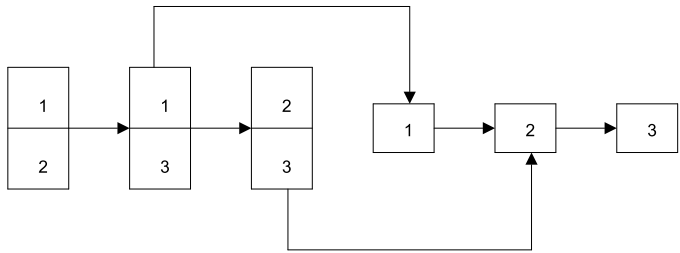} \vskip-17.5cm
\caption{\quad $su(3)$ Hasse diagram with the fundamental
semi-standard tableaux.}
\label{AiiYT} 
\end{center}
\end{minipage}
\end{figure}

The Hasse diagram relevant to $su(3)$ semi-standard tableaux is
drawn in Figure \ref{AiiYT}, with the fundamental semi-standard
tableaux drawn where the corresponding vertices would be. They are
lined up horizontally, to make obvious the connection with the
semi-standard tableaux for $su(3)$. Any number of copies of the
fundamental tableaux of each kind can be used to build a valid
semi-standard tableaux, as long as the partial order is respected.

\begin{figure}[t]
\begin{minipage}{10cm}
\vskip-4cm
\begin{center}
\epsfxsize=17cm \epsfbox{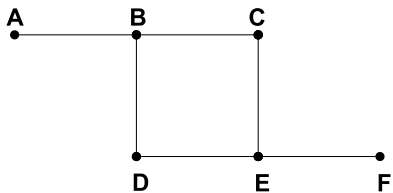} \vskip-15.5cm
\caption{\quad Hasse diagram of the $su(3)$ fundamental poset.}
\label{Aiigraphs} 
\end{center}
\end{minipage}
\end{figure}

\begin{figure}[t]
\begin{minipage}{10cm}
\vskip-0cm
\begin{center}
\epsfxsize=17cm \epsfbox{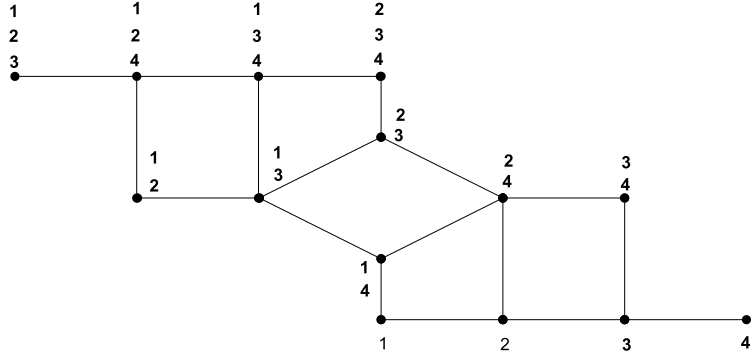} \vskip-17.5cm \caption{\quad
Hasse diagram of the $su(4)$ fundamental poset, with the fundamental
tableaux indicated next to the corresponding vertices.}
\label{Aiiigraph} 
\end{center}
\end{minipage}
\end{figure}

Recall that the weight of an integrity basis element $L^\Lambda\,
a^\mu$, is $\mu$ (while the fundamental weight $\Lambda$ is its
shape). The weights of the fundamental semi-standard tableaux are
the weights of the integrity basis elements (\ref{ABCdefA}) for
the character generator. The Hasse diagram can be labelled by
those basis elements, and then the diagram provides a method of
constructing the generating function directly. For $su(3)$, the
resulting Hasse diagram is the first one drawn in Figure
\ref{Aiigraphs}. The corresponding $su(4)\cong A_3$ Hasse diagram
is shown in Fig. \ref{Aiiigraph}.

Consider the $su(3)$ character generator (\ref{ABCXA}). The two
terms are easily seen to correspond to the two longest paths (or
walks) $ABDEF$ and $ABCEF$ on the Hasse diagram from ``the
beginning'' $A$, or the greatest element, to ``the end'', or least
element, $F$. These two paths correspond to the two maximal chains,
or totally ordered sets, $A\ge B\ge D\ge E\ge F$ and $A\ge B\ge C\ge
E\ge F$ in the corresponding poset. The two maximal chains are
treated in equal fashion, since \ben \lf D \rf\ +\ C\,\lf C \rf\ =\
1\ +\ D\,\lf D \rf\ +\ C\,\lf C \rf\ . \label{CDforb}\een From this
last expression, one can see that the extra factor of $C$ in
(\ref{ABCXA}) is necessary to avoid over-counting.

The point of view just explained was discovered by Baclawski
\cite{Ba}, and applied to the simple algebras $A_r$ and $C_r$ (as
well as to $U(N)$). Using generalized tableaux, character
generators for all classical algebras ($A_r,B_r,C_r$, and $D_r$)
were studied in \cite{KEi, KEii}.

Let us write Baclawski's results, concentrating on the case of
$A_r$. Denote by ${\cal P}$ the poset with fundamental tableaux as
elements, the so-called fundamental poset. We can label the
elements of the poset with the corresponding elements of the
integrity basis $I_\rX$, as in the diagram of Fig. \ref{Aiigraphs}
for $su(3)$. The result is called a labelling of the poset ${\cal
P}$, since the labels can be added and multiplied. A multi-chain
is a chain with repeated elements, such as $m = A\ge A\ge B\ge
D\ge D\ge E\ge F\ge F\ge F$, where, by abuse of notation, we use
the labels to denote the poset elements. The label of such a
multi-chain is easily obtained: \ben \ell\,(m)\ =\ \ell\,(\, A\ge
A\ge B\ge D\ge D\ge E\ge F\ge F\ge F \,)\ =\ A^2BD^2F^3\ .
\label{lmch}\een The first result is simply written as \ben \rX\
=\ \sum_{m\,\in\, M({\cal P})}\ \ell\,(\, m \,)\
,\label{Xsummc}\een where $M({\cal P})$ denotes the set of
multi-chains of ${\cal P}$.

As pointed out above, the relevance to $\rX$ of maximal chains is
immediately obvious. To write the formula \cite{Ba} that makes the
connection explicit, consider the poset $\hat {\cal P}$, the
extended fundamental poset, obtained by adjoining two new
elements, $\hat 0$ and $\hat 1$, to the poset ${\cal P}$. The
element $\hat 0$ satisfies $x\ge \hat 0$, and $\hat 1$ obeys $\hat
1\ge x$, both for all $x\in \hat {\cal P}$. The labelling of $\hat
{\cal P}$ that we use is simply obtained by adjoining the labels
$\ell(\hat 0)= \ell(\hat 1) = 1$ to the labelling of ${\cal P}$.

The links of a poset are relevant here. A chain ${\cal C}$ of a
poset ${\cal P}$ is called saturated if no $z\in {\cal
P}\backslash {\cal C}$ exists such that $x\ge z\ge y$ for $x,y\in
{\cal C}$, such that ${\cal C}\cup\{z\}$ is a chain. Roughly
speaking, there are no gaps in a saturated chain. A cover relation
is a two-element saturated chain, and a link is a saturated chain
with three-elements.

Let ${\rm Link}(\hat {\cal P})$ denote the set of links of $\hat
{\cal P}$. For the $su(3)$ case, we have \bea {\rm Link}(\hat
{\cal P})\ =\ \{\, \hat 1>A>B,\, A>B>C,\, A>B>D,\, B>C>E,\q\nn\qq
B>D>E,\, C>E>F,\, D>E>F,\, E>F>\hat 0 \,\}\ .
\label{LinkPhAii}\eea A linking of a poset ${\cal P}$ is a
partition of ${\rm Link}(\hat {\cal P})$ into two disjoint subsets
${\rm Link}^\pm(\hat {\cal P})$, such that, for every pair $x> y$
in $\hat {\cal P}$, there exists a unique saturated chain $x=x_0>
x_1> \cdots
>x_{n-1}>x_n=y$, every link of which is in ${\rm Link}^+(\hat {\cal P})$.
For the $A_2$ example, one linking of the poset ${\cal P}$ is
specified by the choice \ben {\rm Link}^-(\hat {\cal P})\ =\ \{\,
B>C>E \,\}\ .\label{LinkmAii}\een  Then ${\rm Link}^+(\hat {\cal
P}) = {\rm Link}(\hat {\cal P})\backslash {\rm Link}^-(\hat {\cal
P})$.

Another concept required for the formula is that of a descent set
${\cal DS}(m)$, of a maximal chain $m=x_0>x_1> \cdots > x_n$ of
$\hat {\cal P}$: \ben {\cal DS}(m)\ :=\ \{\, x_i\, |\, 0<i<n\ {\rm
and}\ (x_{i-1}>x_i> x_{i+1})\in {\rm Link}^-(\hat {\cal P}) \,\}\
. \label{Dsetm}\een Its label is therefore \ben \ell\,\left(\,
{\cal DS}(m) \,\right)\ =\ \prod_{x\in \,{\cal DS}(m)}\ \ell\,(\,
x \,)\ .\label{lDsetm}\een Let ${\rm Max}(\hat {\cal P})$ denote
the set of maximal chains in $\hat {\cal P}$. Baclawski \cite{Ba}
proved \ben \rX\ =\ \sum_{m\,\in\, {\rm Max}(\hat {\cal P})}\
\lf\, \ell\,(\, m \,) \,\rf\ \ell\,(\, {\cal DS}(m) \,)\ .
\label{Xsummaxc}\een

For the extended poset $\hat {\cal P}$ relevant to $A_2$, ${\rm
Max}(\hat {\cal P})$ contains two chains, $\hat 1>A>B >D>E>F
>\hat 0$ and $\hat 1>A>B >C>E>F >\hat 0$. For the first, the linking
specified by (\ref{LinkmAii}) gives a null descent set, while for
the second maximal chain, the descent set is $\{ C \}$. The
formula (\ref{Xsummaxc}) therefore immediately reproduces the
result (\ref{ABCXA}).

With the alternate choice ${\rm Link}^-(\hat {\cal P}) = \{ B>D>E
\}$, Baclawski's formula (\ref{Xsummaxc}) yields $X = \lf ABCEF
\rf + \lf AB \rf D \lf DEF\rf$, an equivalent form.

Notice that for both linkings of the poset ${\cal P}$, the product
$CD$ is incompatible. Incompatibilities are fixed by the poset
itself, rather than by a choice of linking of ${\cal P}$. $CD$ is
an incompatible product because $C$ and $D$ are incomparable in
the poset, as is clear from its Hasse diagram in Fig.
\ref{Aiigraphs}.

We should mention that Baclawski \cite{Ba} also derived formulas
of a recursive nature, that lead to nested expressions for $\rX$.
Considerably shortened expressions can result this way, since the
sub-poset structure is taken into account. Since our goal is an
understanding of the full character generator and corresponding
(generalized) posets, however, we will not study those formulas
here.

In Fig. \ref{Aiiigraph} the Hasse diagram of the fundamental poset
for $A_3\cong su(4)$ is depicted, with the vertices labelled by
the corresponding  fundamental semi-standard tableaux. Using
(\ref{Xsummc}) or (\ref{Xsummaxc}) on this diagram yields the
$A_3$ character generator in straightforward fashion. Higher ranks
involve larger fundamental posets and Hasse diagrams, but do not
require any new important complications.

Clearly, the fundamental poset ${\cal P}$ encodes the essence of
the character generator $\rX$, for the algebras $A_r$. This poset
can also be constructed without reference to semi-standard
tableaux. The alternative construction uses the Weyl group and its
Bruhat order (see \cite{HH}, e.g.). That means it is more easily
adaptable to general simple Lie algebras than are the
semi-standard tableaux relevant to $A_r \cong su(r+1)$.

The elements of the poset are in one-to-one correspondence with
the weights of the $r$ fundamental representations of $A_r$. The
poset's cover relations can be stated simply if the vertex
corresponding to the weight $\mu$ in $R(\Lambda^j)$ is indicated
by the triple $[\Lambda^j,\mu; w]$, $w\in W$. That is, we adjoin a
fixed $w\in W$ obeying $\mu=w\Lambda^j$. Of course, there is an
 ambiguity in the choice of $w$ for a fixed  weight $\mu$ of
$R(\Lambda^j)$. Consider, however,
 the reduced decomposition of $w_L$ used in the Demazure calculation of
the character generator.
 For $su(r+1)$, we can use \ben  w_L\ =\ (r_rr_{r-1}\cdots r_1)
 (r_rr_{r-1}\cdots r_2) (r_rr_{r-1}\cdots r_3) \cdots (r_rr_{r-1})(r_r)\ .
 \label{wLsu}\een This expression motivates the label
$[\Lambda^r,\Lambda^r; {\rm id}]$ and,
 for the other highest-weight vertices, \ben [\Lambda^j,\Lambda^j;
 (r_rr_{r-1}\cdots r_{j+1}) \cdots (r_rr_{r-1})(r_r) ]\ ,
 \label{hwvw}\een for $j=1,\ldots,r-1$. Notice that
 the length of the adjoined Weyl elements increases as $j$ decreases in
 $\Lambda^j$. That is, our choice of reduced decomposition for
 $w_L$ induces a total order on the set $F$ of fundamental
 weights.

\begin{figure}[t]
\begin{minipage}{10cm}
\vskip-0cm
\begin{center}
\epsfxsize=17cm \epsfbox{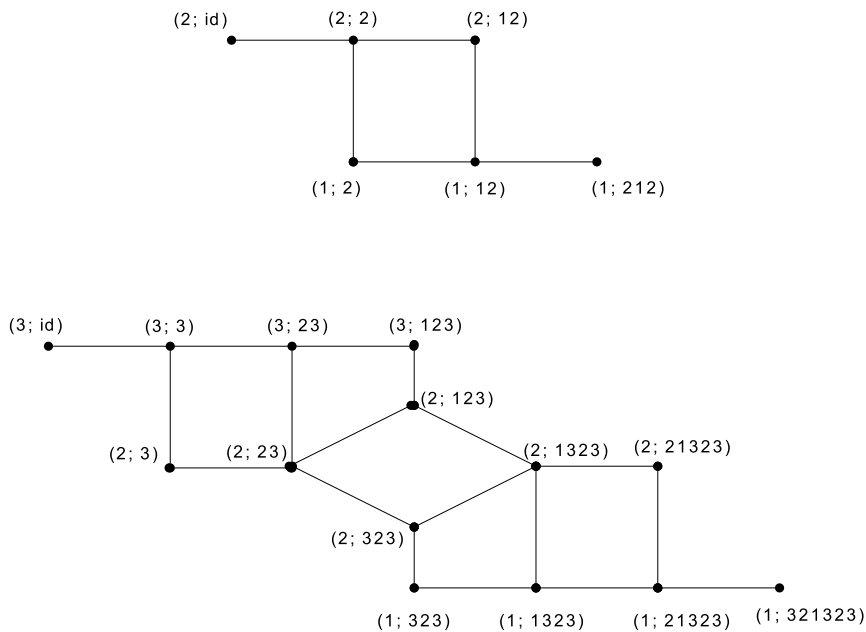} \vskip-13.5cm \caption{\quad
Hasse diagrams for the $su(3)$ and $su(4)$ fundamental posets.
Next to the vertices, the corresponding integrity basis elements
are indicated, by their shapes and a Weyl group element that maps
their shape to their weight. For example, (2; 123) indicates shape
$\Lambda^2$ and weight $r_1r_2r_3\Lambda^2 =
-\Lambda^1+\Lambda^3$, i.e., $L_2a_1^{-1}a_3$.}
\label{Aiiigraphs} 
\end{center}
\end{minipage}
\end{figure}

 Once the Weyl elements are fixed for the highest-weight vertices, the Bruhat
 order can then be used to assign Weyl elements to all the
 vertices of the required Hasse diagram. The edges of the Hasse
 diagram are determined by the cover relations
 \bea [\Lambda^j, \mu; w]\
\rightarrow\ [\Lambda^i, \nu; v]\, , \q {\rm if}\ i=j\ {\rm and}\
w\leftarrow v\, ,\q\nn \QQ\ {\rm or\ if}\ j=i+1\ {\rm and}\ w=v\
.\label{iwcover}\eea Here $w\leftarrow v$ indicates a cover
relation in the Bruhat partial order on $W$. See Fig.
\ref{Aiiigraphs} for illustrations of the cases $A_2$ and $A_3$.

Before leaving the $A_r$ algebras, we should mention that the
semi-standard tableaux have significance for the vectors (states)
in representations, not just their multiplicities. As is well
known, a vector in an arbitrary irreducible representation of
$A_r$ can be constructed from the vectors of the fundamental
representations, which are in turn constructible from the vectors
of the first fundamental (basic) representation. The latter can be
labelled by single boxes, numbered from 1 to $r+1$. Totally
antisymmetric $j$-fold tensor products of the basic vectors yield
the vectors of the fundamental representation of highest weight
$\Lambda^j$. Symmetrizing these, according to the rows of a fixed
Young tableau, produces the vectors of the representation of
highest weight equal to the tableau shape.

Consequently, the generating function $\rX$ and the related
fundamental poset ${\cal P}$ encode something of this
construction.\footnote{\ Arguably, the most important use of the
character generator is to tell us about this method of building
vectors of highest-weight representations.} Conversely, knowing
how to construct the vectors from those of the fundamental
representations, can tell us about the character generator $\rX$.
As we will discuss below, the $G_2$ character generator was found
this way in \cite{BT}.

\subsection{$B_2$}

The $A_r$ case is simple. All fundamental representations are
minuscule, i.e., their weights form a single Weyl orbit
$W\Lambda^j$. This means, in particular, that there are no inside
generators in the case of $A_r$.

For the algebra $B_2$, however, there is one inside generator, $z\
:=\ \Db_2L_1a_1^{-1}a_2^{2}$ $=\ L_1$. From the expression
(\ref{ABCXB}), it is clear that this inside generator $z$ is not
treated in the same way as the outer generators $A,\ldots,H$.
While $\rX$ is linear in $z$, it contains arbitrarily high powers
of each of the outside generators.

\begin{figure}[t]
\begin{minipage}{10cm}
\vskip-3cm
\begin{center}
\epsfxsize=17cm \epsfbox{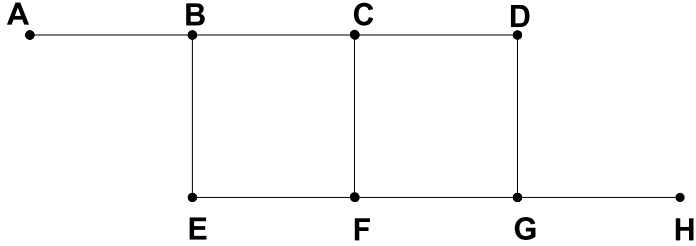} \vskip-16.5cm
\caption{\quad  Hasse diagram of the fundamental-orbit poset of
$B_2$.}
\label{Biigraph} 
\end{center}
\end{minipage}
\end{figure}

As in the $A_2$ case, however, the result (\ref{ABCXB}) can be
understood in terms of a graph related to a poset. We can use the
same construction as for $A_2$, including the cover relations
(\ref{iwcover}), as long as only the elements of the Weyl orbits
$W\Lambda^j$ are included in the poset. For the $B_2$ case, the
Hasse diagram of this poset is drawn in Fig. \ref{Biigraph}. We
will call the poset so constructed the fundamental-orbit poset
${\cal P}_o$.

The connection of (\ref{ABCXB}) to ${\cal P}_o$ is clear. If we
modify the labelling of the maximal chains of ${\cal P}_o$ so that
\ben \ell\,(\, \cdots > F > G > \cdots \, )\ =\ \cdots\,F\,(1+z)\,
G\, \cdots\ \ \ , \label{modlB}\een then Baclawski's formula
(\ref{Xsummaxc}) still works. That is, in maximal chains, we
introduce a labelling of the edge $\{F,G\}$ of the Hasse diagram
connecting vertices $F$ and $G$, that corresponds to the cover
relation $F>G$.

The extra labelling has a Demazure interpretation: $z = \Db_2F$
and \ben \hD_2\,\lf F \rf\ =\ \lf F\rf\, \left(\, 1 +  \Db_2 F
\,\right)\,\lf \hr_i F \rf\ =\ \lf F\rf\, \left(\, 1 +  z
\,\right)\,\lf G \rf\ . \label{DFGdmr}\een by eqn.
(\ref{DisFdmr}).

The latter result shows that only the $\{F,G\}$ edge needs this
extra factor, because no outer generator other than $F$ has a
weight with a Dynkin label greater than 1. If generator $V$ has
$i$-th Dynkin label equal to 1, then $\Db_i V = 0$. We can,
therefore extend the labelling to include all the edges of the
Hasse diagram between vertices with weights in the same
fundamental representation. Label with $(1 + \Db_jV)$ the edge
$\{V,\hr_j V\}$. For all cases considered so far, except $V=F$,
this label is just 1.

\subsection{$G_2$}

For $G_2$, $\Vert I_{\rm in} \Vert = 9$, so the situation becomes
more complicated. That is made plain by looking at the final
expression for $\rX$. The fundamental-orbit poset ${\cal P}_o$ has
$\Vert I_{\rm out} \Vert = 12$ elements, and is easily
constructed. If, as for $B_2$, we continue to label with $(1 +
\Db_jV)$ the edge $\{V,\hr_j V\}$, there are many terms
recognizable in (\ref{AGiiXDf}) as coming from the maximal chains
of ${\cal P}_o$. However, there are many more terms in
(\ref{AGiiXDf}) than there are maximal chains in ${\cal P}_o$.

\begin{figure}[t]
\begin{minipage}{10cm}
\vskip-4cm
\begin{center}
\epsfxsize=20cm \epsfbox{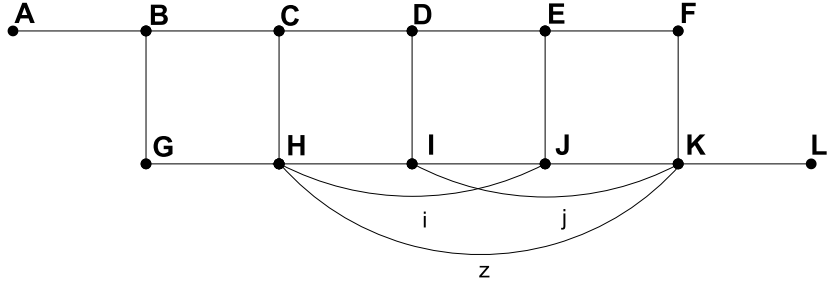} \vskip-17.5cm \caption{\quad
$G_2$ graph.}
\label{OWgraph} 
\end{center}
\end{minipage}
\end{figure}

To proceed, we introduce new edges to the Hasse diagram of ${\cal
P}_o$, and label the new edges as needed to produce the expression
(\ref{AGiiXDf}). The result is the graph of Figure \ref{OWgraph},
where only the labels of the new edges are indicated. Three new
edges are required: the $\{H,J\}$ edge with label $\hd_1\Db_2H =
i$, $\{I,K\}$ with label $\hr_2\hd_1\Db_2H = j$, and $\{H,K\}$
labelled by $\Db_2\hd_1\Db_2H = z$. Consider now maximal paths (or
walks) on this graph, from the beginning vertex $A$ to the end
$F$. The terms of (\ref{AGiiXDf}) can be put in one-to-one
correspondence with the maximal paths, so that a formula of the
Baclawski type can still be written, as long as the edge labels
are included as factors.

The final element required is an explanation of the edge labels.
They are sums of inside generators, but which ones? Their
individual expressions in terms of Demazure operators are not
particularly illuminating.

However, notice that the new edges only relate outside generators
with weights from the same fundamental representation, say
$R(\Lambda)$, for some $\Lambda\in F$. Focus on a vertex
$L^{\Lambda}a^{v\Lambda} = \hat v\,( L^{\Lambda}a^{\Lambda})$,
where $v\in W$. It is clear that any inside generators with
weights in $\hd_v a^{\Lambda}$ must appear as labels of edges
ending on that vertex.

Consider the label $\hd_1\Db_2H = i$ of the new $\{H,J\}$ edge.
Since $J = \hr_1\hr_2\hr_1 G$, we calculate \ben
\hd_{r_1r_2r_1}\,G\ =\ \hd_1\, \hd_2\, \hd_1\, G\ =:\
\hd_{1\,2\,1}\, G\ =\ i\ +\ z'\ +\ J\ . \label{dGJ}\een The inside
generator $i$ labels one edge ending on $J$, while $(1+z')$ is the
label of the other edge ending there.

This way, the labels of edges ending on a graph vertex can be
found. To determine where the edges should begin, we simply
reverse the process. Start with the outside generator of lowest
weight, $L^{\Lambda}a^{w_L\Lambda} = \hat w_L\,(
L^{\Lambda}a^{\Lambda})$, $\Lambda\in F$. To work backwards, we
need to consider Demazure operators like the $\hd_m$, but where
the role of the simple root $\alpha_m$ is taken by its negative
$-\alpha_m$. We denote such a Demazure operator by
$\hd_{\underline{m}}$, and also use the convention that
$\hd_{\,{\underline \ell}\, {\underline m}} :=
\hd_{\underline\ell}\, \hd_{\underline m}$, etc.

For the $\{H,J\}$ edge, the generator of lowest weight is $L$. We
calculate \ben  \hd_{\,\underline{2}\,
\underline{1}\,\underline{2}\, \underline{1}}\,L\ =\
\hd_{\underline{2}}\, \hd_{\underline{1}}\, \hd_{\underline{2}}\,
\hd_{\underline{1}}\, L\ =\ z + i + h + g + H\ . \label{bdLH}\een
Comparing (\ref{bdLH}) and (\ref{dGJ}), we see that only $i$ is
common, and so $i$ will label the edge beginning at $H$ and ending
on $J$.

This procedure works for all the new edges. It is easy to find
\bea \hd_{\,2\, 1\, 2\,1}\, G\ =\ z+j+k+\ell+K\ , \qq\nn
\hd_{\underline{1}\,\underline{2}\,\underline{1}}\, L\ =\ j+z'+I\,
,\ \ \hd_{\underline{2}\,\underline{1}\,\underline{2}\,
\underline{1}}\, L\ =\ z+i+h+g+H\ .\label{newdGdbL}\eea The first
two of these results confirms that the $\{I,K\}$ edge is labelled
by $j$; the first and third give $z$ as the $\{H,K\}$ label.

The nontrivial $\Db_i V$ part of the labels $(1+\Db_i V)$ for the
edges $\{V,\,\hr_iV\}$ of the Hasse diagram of ${\cal P}_o$ that
is contained in our graph, can also be obtained this way.

\begin{figure}[t]
\begin{minipage}{10cm}
\vskip-9cm
\begin{center}
\epsfxsize=20cm \epsfbox{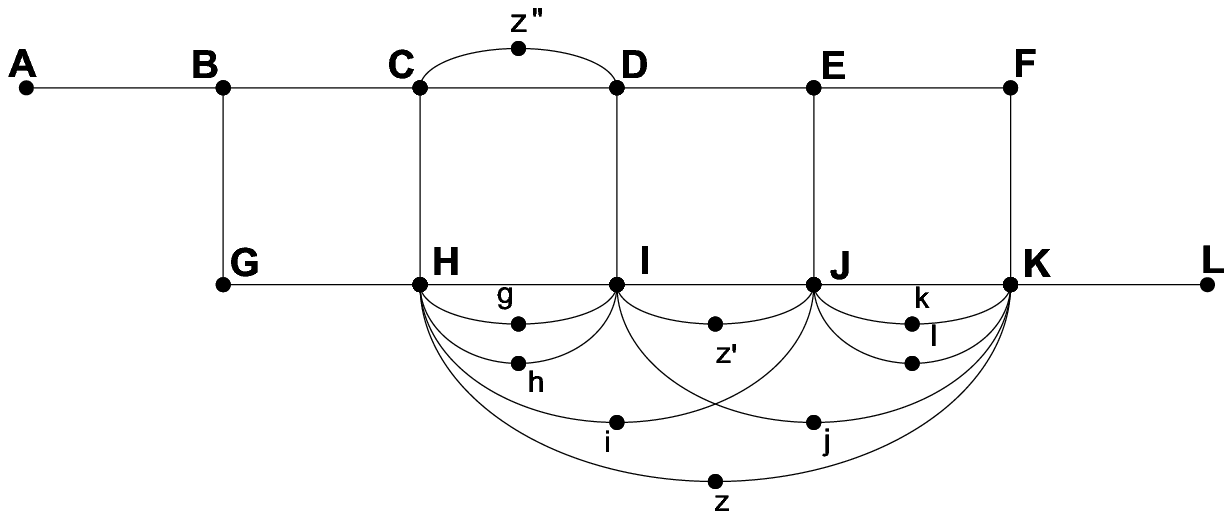} \vskip-11.5cm
\caption{\quad The $G_2$ generalized-poset graph of Baclawski and
Towber \cite{BT}.}
\label{BTgraph} 
\end{center}
\end{minipage}
\end{figure}

Our graph, as shown in Fig. \ref{OWgraph}, has an obvious
resemblance to that devised long ago by Baclawski and Towber
\cite{BT}, depicted in Fig. \ref{BTgraph}. Every element of the
integrity basis (both inside and outside generators) is
represented by a vertex in that graph. The Baclawski-Towber graph
is not the Hasse diagram of a poset, however, but rather its
generalization for a generalized poset. The generalization is
necessary because an inner generator $\iota$ appears at most
linearly in $\rX$. That means $\iota^2$ is an incompatible
product. Incompatibilities correspond to incomparable elements of
a poset, however, and the poset partial order $\ge$ obeys the
reflexivity property: $x\ge x$, for all $x$ in a poset ${\cal P}$.
For $x$ in a poset, then, $x$ is always comparable to $x$.

If the partial order $\ge$ is replaced by a binary relation $\gg$
without reflexivity, however, a generator $\iota$ can be
incomparable to itself. All the inner generators $\iota$ do not obey
$\iota\gg \iota$, and so $\iota^2$ can be an incompatible product.
In the generalized poset, the inner generators are incompatible with
themselves, while the outer generators are not. As a consequence,
the vertices of the corresponding graph are not all treated on an
equal footing. With this modification, however, formulas like the
poset ones written by Baclawski  \cite{Ba} can also be written for
graphs of the type in Fig. \ref{BTgraph}.

In contrast, we prefer to work with a graph more closely related
to the Hasse diagram of the fundamental-orbit poset ${\cal P}_o$.
We do not increase the number of vertices by introducing new ones
for every inner generator, i.e., for every integrity basis element
with a weight not an element of a fundamental Weyl orbit
$W\Lambda$, $\Lambda\in F$. Instead, we introduce edge labels
involving the inner generators for the edges of the Hasse diagram
of ${\cal P}_o$, and add new edges (only) with such labels. In our
opinion, the resulting graph is simpler than that of ref.
\cite{BT}; compare Figs. \ref{OWgraph} and \ref{BTgraph}. We will
call our graph and its generalization to other simple Lie algebras
the character-generator graph, and denote it ${\cal G}_\rX$.

We should point out, however, that just as Baclawski and Towber
treat the vertices for inner and outer generators differently, we
do not treat all the edges of ${\cal G}_\rX$ on equal footing. An
edge between two outer generators related by a primitive
reflection $r_j$ gets special treatment. The 1 of the labels
$(1+\Db_i V)$ of the edges $\{V,\hr_iV\}$ must be added.

If we focus on the character generator of the algebra $G_2$ only,
our result just amounts to a slight simplification of that of
\cite{BT}. On the other hand, an important difference is revealed
if we compare the methods used.

As was discussed above, semi-standard tableaux reveal the poset
structure underlying the $su(r+1)\cong A_r$ character generator.
They also encode a construction of the vectors of an irreducible
highest-weight representation, using as a basis the vectors of the
fundamental representations. While writing down the vectors of a
fixed representation is much more involved than finding its
weights and multiplicities, doing the former does tell us about
the latter, and so about the character of the representation. A
general construction, for all highest weight representations can,
therefore, reveal the structure of the character generator.

This construction of vectors is possible for any simple Lie
algebra, providing a way to the character generator of that
algebra. In \cite{BT}, the authors defined what they called a
shape algebra, which is useful for such constructions, but is
framed in a more general context. More importantly for us, they
constructed the required basis for $G_2$ explicitly, and were
consequently able to draw the generalized poset graph of Fig.
\ref{BTgraph}, and write the character generator $\rX$. Their work
used the special relation of $G_2$ to the octonions $\O$,
however.\footnote{\ The $G_2$ algebra is the algebra of
derivations of $\O$. It is true that all the
$A_r,B_r,C_r,D_r,E_6,E_7,E_8,F_4$ algebras can be related in a
similar way to the four normed division algebras: $\R,\C,$ the
quaternions $\H$, and the octonions $\O$ (see \cite{Bae}, e.g.).
Even so, $G_2$ does not fit nicely into the pattern filled out by
the others. For example, the $E$ and $F$ exceptional algebras are
elements of the so-called magic square, while $G_2$ is not.} It
was therefore not able to yield results in a general form, useful
for any simple Lie algebra. Their $G_2$ results were not derived
or written in terms of objects common to all simple Lie groups
and/or their algebras, like the Weyl group, for example.

Our method, however, is essentially that of Gaskell \cite{G},
taking into account the poset structure of Baclawski \cite{Ba}. As
such it uses only general methods, involving Weyl groups and their
Bruhat order, Demazure operators, and a total ordering of
fundamental weights induced by a reduced decomposition of $w_L$.
It therefore leads to results that we believe indicate the general
form of the character generator for all simple Lie algebras.

\subsection{General simple Lie algebras - possible universal picture}

Let us now sketch the construction of the character-generator
graph ${\cal G}_\rX$ in terms that apply to an arbitrary simple
Lie algebra $X_r$.

We must emphasize that the construction outlined below has not
been proven correct. All we can say at this point is that it works
for the algebras we considered, and is expressed in universal
terms. It therefore has a hope of applying to all simple Lie
algebras.

First, construct the fundamental-orbit poset ${\cal P}_o$. Its
elements are in one-to-one correspondence with the outside
generators of the integrity basis (\ref{Xbas}) for $\rX$, and so
with pairs $(\Lambda,\mu)$, where $\Lambda\in F$ and $\mu\in
P_\Lambda$, the set of weights (of non-zero multiplicity) in the
fundamental representation $R(\Lambda)$. We write \ben {\cal P}_o\
=\ \{\ [\Lambda,\mu;w]\ |\ \Lambda\in F, \mu\in P_\Lambda; w\in
W{\rm\ such\ that\ }\mu=w\Lambda\ \}\ .\een Notice that only one
Weyl group element is associated with each pair $(\Lambda,\mu)$,
i.e. each element of ${\cal P}_o$. The choice of these Weyl
elements is not unique; the different possible choices allow the
same character generator $\rX$ to be described by different
posets.

To make one such choice, fix a reduced decomposition of $w_L$, and
write it as \ben w_L\ :=\ s_Ls_{L-1}\cdots s_1\ , \label{wLrd}\een
where each $s_a$ is a primitive reflection of $W$, so that
$L=\ell(w_L)$, the length of $w_L$. More generally, we will use
\ben w_{L,a}\ :=\ s_as_{a-1}\cdots s_1\ .\label{ward}\een Set the
highest-weight elements of ${\cal P}_o$ to be
$[\Lambda^j,\Lambda^j; w_{\rm max}^{(j)}]$, where $w_{\rm
max}^{(j)}$ is the longest of the Weyl group elements $w_{L,a}$
fixing $\Lambda^j$: $ w_{\rm max}^{(j)}\Lambda^j = \Lambda^j$.
Then the remaining elements can be assigned Weyl group elements
using the Bruhat order: $[\Lambda, \mu; w]\
>\ [\Lambda, \nu; v]$ if $w\prec v$.

The reduced decomposition of $w_L$ selected also induces a total
order $\ge$ on the fundamental weights of $F$. Let $\Rightarrow$
denote its cover relations. We put $\Lambda^j > \Lambda^i$ if
$\ell(w_{\rm max}^{(j)}) < \ell(w_{\rm max}^{(i)})$.

The partial order of ${\cal P}_o$ can then finally be fully
defined by the cover relations \bea [\Lambda, \mu; w]\
\rightarrow\ [\Lambda', \nu; v]\, , \quad{\rm if}\
\Lambda=\Lambda'\ {\rm and}\ w\leftarrow v\, ,\qq\nn \qq\ \ \ {\rm
or\ if}\ \Lambda\Rightarrow\Lambda'\ {\rm and}\ w=v\ .\q
\label{porderPo}\eea

Let ${\rm E}({\mathcal{G}})$ and ${\rm V}({\mathcal{G}})$ indicate
the edge set and the vertex set, respectively, of a graph
${\mathcal{G}}$. The character-generator graph ${\mathcal{G}}_\rX$
is built on the skeleton ${\mathcal{H}}({\cal P}_o)$. More
precisely, \ben V\left({\mathcal{G}}_\rX\right)\ =\  V\left(
{\mathcal{H}}({\cal P}_o) \right)\ ,\ \ \
E\left({\mathcal{G}}_\rX\right)\  \supset\  E\left(
{\mathcal{H}}({\cal P}_o) \right)\ .\label{GXHPo}\een

Label the vertices of the Hasse diagram ${\mathcal{H}}({\cal
P}_o)$ of ${\cal P}_o$ using \ben \ell\left(\, [\Lambda, \mu; w]
\,\right)\ =\ L^\Lambda\, a^{\mu}\ . \label{ellvPo}\een We will
also label the edges of the resulting character-generator graph,
using Demazure objects. First, all edges of the ${\cal P}_o$ Hasse
diagram are labelled by $1$. Additional labels are introduced as
follows, and they will add to the $1$s already present, or label
new edges of ${\mathcal{G}}_X \supset {\mathcal{H}}({\cal P}_o)$,
when they do not vanish.

The edge labels and the ``new'' edges, the elements of ${\rm
E}({\cal G}_\rX)\backslash{\rm E}\left( {\cal H}({\cal P}_o)
\right)$, are found using Demazure calculations. Suppose that
$T=L^\Lambda$ and $B=\hat w_LT$ indicate the top and bottom
vertices of the same shape $\Lambda\in F$. Consider the vertices
$V_1$ and $V_2$, with $V_1>V_2$ in ${\cal H}({\cal P}_o)$. Suppose
further that $V_1=\hat b_1B$, and $V_2= \hat t_2T$, with
$b_1,t_2\in W$. Calculate $\hd_{\,t_2}\,T$ and
$\hd_{\,\underline{b_1}}\,B$. Denote the sum of terms common to
both as \ben  d(V_1,V_2)\ :=\ \hd_{\,\underline{b_1}}\,B\ \cap\
\hd_{\,t_2}\,T\ .\label{dintd}\een  If $d(V_1,V_2)\not= 0$,  then
$\{V_1,V_2\}$ will belong to ${\rm E}({\cal G}_\rX)$.

The labels of the edges of ${\cal G}_\rX$ are given by  \ben
\tilde\ell\,\big(\, \{V_1,\, V_2\} \,\big) =\ \left\{\matrix{1\ +\
d(V_1,V_2)\ , &\ {\rm if}\ \ \{V_1,V_2\}\, \in\, {\rm E}({\cal
H}({\cal P}_o))\ ;\cr d(V_1,V_2)\ , &\ {\rm if}\ \ \{V_1,V_2\}\,
\not\in\, {\rm E}({\cal H}({\cal P}_o))\
.}\right.\label{elledge}\een

We can write a formula analogous to (\ref{Xsummaxc}) for the
general case if we consider ${\cal G}_\rX$ the Hasse diagram of a
poset ${\cal P}_\rX$. The new poset ${\cal P}_\rX$ has the same
elements as the fundamental-orbit poset ${\cal P}_o$, but its
cover relations are those of ${\cal P}_o$ augmented by those
encoded in the new edges of ${\cal G}_\rX$.

Maximal chains in ${\cal P}_\rX$ are relevant here, but their
labels must include the edge labels as factors, along with those
of the vertices. We define \ben \tilde\ell\,( \cdots V_1
>\, V_2 \cdots )\ :=\ \cdots\,\tilde\ell\,(V_1)\,
\tilde\ell\,(\{V_1,V_2\})\, \tilde\ell\,(V_2)\, \cdots \
.\label{labelPX}\een The symbol $\tilde\ell$ indicates the
labelling of ${\cal G}_\rX$, to distinguish it from the labelling
$\ell$ of ${\cal P}_o$.

We need the extended poset  $\hat {\cal P}_\rX$, and its
labelling. But its labelling is trivially different from that of
${\cal P}_\rX$: vertices $\hat 1$ and $\hat 0$, and the two extra
edges involving them, are all assigned 1 as labels. We will also
use $\tilde\ell$ for the labels of $\hat {\cal P}_\rX$.

Incompatible products are treated in (\ref{Xsummaxc}) using
linkings of the extended poset $\hat {\cal P}$ and the resulting
descent sets, defined in (\ref{Dsetm}). In the general case, only
incompatibilities between two outside generators need to be
handled this way. Therefore, it is the linking of $\hat {\cal
P}_o$ that is relevant. Suppose $m$ is a maximal chain in $\hat
{\cal P}_\rX$. Then we define \ben {\cal DS}_o(m)\ :=\ \{\, x_i\,
|\, 0<i<n\ {\rm and}\ (x_{i-1}>x_i> x_{i+1})\in {\rm Link}^-(\hat
{\cal P}_o) \,\}\ . \label{tDsetm}\een

Finally, we are able to write \ben \rX\ =\ \sum_{m\,\in\, {\rm
Max}(\hat {\cal P}_\rX)}\ \lf\, \tilde\ell\,(\, m \,) \,\rf\
\ell\,(\, {\cal DS}_o(m) \,)\ . \label{XsummaxPXc}\een Here the
shorthand notation of (\ref{floor}) and (\ref{bABb}) only applies
to the vertex factors: \ben \lf\, \tilde\ell\,( \cdots V_1
>\, V_2 \cdots ) \,\rf\ :=\ \cdots\, \lf\,\tilde\ell\,(V_1) \,\rf \,
\tilde\ell\,(\{V_1,V_2\})\, \lf\,\tilde\ell\,(V_2) \,\rf\, \cdots
\ .\label{flabelPX}\een

The formula (\ref{XsummaxPXc}) for the character generator $\rX$
is one of our main results. Hopefully, our conjecture generalizes
Baclawski's formula (\ref{Xsummaxc}) so that it can be applied to
any simple Lie algebra.

\vskip1cm\section{Conclusion}

Let us first summarize our main results.

A new, universal formula was derived for the character generator
of a simple Lie algebra. The character generator $\rX$ is
expressed as the ratio $\rX= \rY/\rZ$, with the simple denominator
given by (\ref{XYZ}), and the numerator by (\ref{Yiii}), or,
equivalently, by (\ref{YywK}) and (\ref{K}). The new formula does
not involve a sum over the Weyl group, and so is a simplification
of the Patera-Sharp formula. It a also makes clear the distinct
roles of the inside and outside generators, and can serve as a
guide to incompatible products, as Sect. 3 indicates.

In the second part of this paper, the general, Demazure methods of
Gaskell \cite{Ga} were exploited, and connected with the
(generalized-)posets underlying the character generator
\cite{S,Ba,K,KEi,KEii,BT}. Graphs were found that are simplified
versions of those introduced by Baclawski-Towber \cite{BT}, from
which the character generators can be determined easily. In
particular, the required labelling of the edges of these graphs is
given by simple Demazure calculations. By combining the universal
Demazure-Gaskell approach with the graph structure of the
character generators, we were able to formulate a general
conjecture, that we hope is applicable to all simple Lie algebras.
Thus, non-negativity did not have to be sacrificed to attain
universality. The general formula is eqn. (\ref{XsummaxPXc}), and
it is explained in the rest of subsect. 5.4.

Our second main result is only a conjecture, and clearly lacks
rigour. Possible future work therefore includes proving
(\ref{XsummaxPXc}). Induction may be helpful, and one might be
able to extend our result to a generalization of the character
generator, \ben \rX_w\ :=\ \hD_w\, \rH\ ,\ \ \ \ w\in W\ .
\label{rXw}\een Here $\rX_{w_L} = \rX$.

Alternatively, the general character formulas of Littelmann,
written in terms of minimal defining chains \cite{L} and
Lakshmibai-Seshadri paths \cite{Li}, could provide another route
to the character generators, and a proof.

The character-generator formulas could also be investigated to see
what they tell us directly about the characters themselves. Can a
new character formula be written? Can one derive new identities
involving the characters? The relation of the character generators
and the corresponding integrity bases to bases of states (or
vectors) in irreducible highest-weight representations should also
be understood.

There is a fundamental correspondence found by C. Greene (see
\cite{BF} for a review) that associates to every finite poset a
Young tableau, or Ferrers shape. It would be interesting to try to
apply the correspondence, or a modified version thereof, to the
(generalized-)posets underlying the character generators. A
significantly more economical presentation of the character
generators might result. We suspect that in the simplest cases,
the early results of Stanley \cite{S} and King \cite{K} would be
recovered.

Let us conclude by describing an application of character
generators that was the original motivation for this work.
Two-dimensional conformal field theories \cite{DMS} have been
intensely investigated for quite some time now. Important
examples, the Wess-Zumino-Witten models, are intimately related to
simple Lie algebras. Their so-called modular data (see
\cite{DMS,G,GW}), including their fusion eigenvalues, are
fundamental characteristics. But the fusion eigenvalues of
Wess-Zumino-Witten models coincide with the characters of simple
Lie algebras, evaluated at certain finite-order elements of the
corresponding Lie group. Thus the character generator of a simple
Lie algebra can be used to study Wess-Zumino-Witten fusion
eigenvalues. One of us (M.W.) hopes to make progress in this
direction. The work \cite{MPS} studied character generators for
elements of finite order and so should be helpful.

\vskip1cm
\appendix{\noindent\large{\bf Appendix}\ \  Alternate form of the
Weyl character formula and an identity} 

The identity ({\ref{chO}) can be seen most easily using a
different form of the Weyl character formula: \begin{equation}
{\rm ch}_\lambda\ =\ \sum_{w\in W}\, a^{w\lambda}\,
\prod_{\alpha\in \Delta_+}\, (1-a^{-w\alpha})^{-1}\ .
\label{Weyli}\end{equation} The usual formula (\ref{wcf}) is
recovered from this as follows. Each Weyl element $w\in W$
separates the positive roots into two disjoint sets: \bea
\Delta_+^w:= \{\alpha\in \Delta_+\,|\, w\alpha\in \Delta_+\}\ ,&\
\ \Delta_-^w:= \{\alpha\in \Delta_+\,|\, w\alpha\in \Delta_-\}\
,\cr \Delta_+^w \cup \Delta_-^w\ =\ \Delta_+\ ,&\ \ \Delta_+^w
\cap \Delta_-^w\ =\ \{\}\ ,\cr w\Delta_+^w\ =\ \Delta_+^w\ \ \ \
,&\ \ \ \ w\Delta_-^w\ =\ -\Delta^w_-\ \ . \eea It can be shown
that $\det w = (-1)^{\Vert \Delta^w_-\Vert}$, and
\label{wcfii}\begin{equation} -w\rho +\rho\ =\ \sum_{\beta\in
\Delta^w_-}\, \beta\ \ .
\end{equation}
Using these results, (\ref{Weyli}) becomes \bea {\rm ch}_\lambda\
=\ \sum_{w\in W}\, a^{w\lambda}\, \prod_{\beta\in \Delta^w_-}\,
(-a^{w\beta})(1-a^{w\beta})^{-1}\, \prod_{\alpha\in \Delta^w_+}\,
(1-a^{-w\alpha})^{-1}\ \qq\cr  =\ \sum_{w\in W}\,(\det w)\,
a^{w\lambda-w\sum_{\gamma\in \Delta^w_-}\gamma}\, \prod_{\beta\in
\Delta^w_-}\, (1-a^{w\beta})^{-1}\, \prod_{\alpha\in \Delta^w_+}\,
(1-a^{-w\alpha})^{-1}\ , \eea so that (\ref{wcf}) results.

Now, (\ref{Weyli}) gives \begin{equation} {\hch}\ =\ \sum_{w\in
W}\, \hat w\, \prod_{\alpha\in \Delta_+}\, (1-a^{-\alpha})^{-1}\ .
\label{chWeyli}\end{equation} Therefore, \ben \hch\, \left(\,
a^\mu\, {\cal O}_\lambda (a) \,\right)\ =\  c_\lambda\, \sum_{w\in
W}\, a^{w\mu} \, \prod_{\alpha\in \Delta_+}\,
(1-a^{-w\alpha})^{-1}\, \sum_{u\in W}\, a^{wu\lambda}\ .
\label{usum}\een After a simple change of summation variables, we
find ({\ref{chO}).

\vskip1cm \noindent{\bf Acknowledgments}\hfill\break \noindent
This research was supported by NSERC of Canada. We thank the
Department of Applied Mathematics at the University of Western
Ontario for its hospitality. For their comments, we are grateful
to Chris Cummins, Pierre Mathieu, and especially to Terry Gannon
and Dave Morris. M.W. also thanks the Perimeter Institute, where
some of this work was done.



\begin{thebibliography}{99}

\bibitem{Ba} K. Baclawski, J. Math. Phys. {\bf 24} (1983) 1688

\bibitem{BT} K. Baclawski, J. Towber, Amer. J. Math. {\bf 106}
(1984) 1107

\bibitem{Bae} J. Baez, Bull. Amer. Math. Soc. {\bf 39} (2002) 145

\bibitem{BF} T. Britz, S. Fomin, Finite posets and Ferrers shapes,
e-print arXiv:math.CO/9912126, (1999)

\bibitem{CCS} M. Couture, C. Cummins, R.T. Sharp,  J. Phys. A
{\bf 23} (1990) 1929

\bibitem{D} M. Demazure, Ann. Ecole Norm. Sup. {\bf 7} (1974) 53;
Bull. Sci. Math. 2 Ser. 2 {\bf 98} (1974) 163

\bibitem{DMS} P. Di Francesco, P. Mathieu, D. S\'en\'echal,
Conformal Field Theory (Springer-Verlag, 1997)

\bibitem{FH} W. Fulton, J. Harris, Representation Theory
(Springer-Verlag, 1991)

\bibitem{G}  T. Gannon,  J. Algebraic Combin.  {\bf 22} (2005) 211

\bibitem{GW} T. Gannon, M.A. Walton, Commun. Math. Phys. {\bf 206} (1999) 1

\bibitem{Ga} R. Gaskell,  J. Math. Phys. {\bf 24} (1983) 2379

\bibitem{GS} R. Gaskell, R.T. Sharp,  J. Math. Phys. {\bf
22} (1981) 2736

\bibitem{HS} N. Hambli, R.T. Sharp,
Advances in Mathematical Sciences: CRM's 25 years (Montreal,
1994), 415-419, CRM Proc. Lecture Notes {\bf 11}  (AMS, 1997)

\bibitem{HH} H. Hiller, Geometry of Coxeter Groups (Pitman,
1982); J.E. Humphreys, Reflection Groups and Coxeter Groups
(Cambridge, 1990)

\bibitem{K} R.C. King, C. R.
Math. Rep. Acad. Sci. Canada {\bf 3} (1981) 149

\bibitem{KEi} R.C. King, N.G.I. El-Sharkaway, C. R. Math. Rep. Acad. Sci.
Canada {\bf 4} (1982) 299

\bibitem{KEii} R.C. King, N.G.I. El-Sharkaway, J.
Phys. A {\bf 17} (1984) 19

\bibitem{L} P. Littelmann, J. Alg. {\bf 130} (1990) 328

\bibitem{Li} P. Littelmann, Invent. Math. {\bf 116} (1994) 329

\bibitem{MPS} R.V. Moody, J. Patera, R.T. Sharp, J. Math. Phys. {\bf
24} (1983) 2387

\bibitem{PS} J. Patera, R.T. Sharp,  Proc.
Group Theoretical Methods in Physics (Austin, 1978), Lecture Notes
in Physics {\bf 94}, 175-183 (Springer-Verlag, 1979)

\bibitem{P} J. Patera, Symmetry in physics -- in memory of
Robert T. Sharp, edited by P. Winternitz et al, CRM Proceedings \&
Lecture Notes, vol. 34 (AMS,  2004) 159


\bibitem{S} R.P. Stanley,  J.
Math. Phys. {\bf 21} (1980) 2321

\bibitem{St} R.P. Stanley, Enumerative Combinatorics, Volume 1
(Cambridge Univ. Press, 1997)




















\end{thebibliography}
\end{document}